\documentclass[prx,twocolumn,english,superscriptaddress,floatfix,longbibliography]{revtex4-2}
\usepackage{multibib}
\usepackage{booktabs}
\usepackage{times}
\usepackage{braket}
\usepackage{mathptmx}
\usepackage{cancel}
\usepackage{adjustbox}
\usepackage{amsmath, amsthm, thm-restate, bbold, physics, amssymb}
\usepackage{braket, enumitem, relsize}
\usepackage{makecell, graphicx, subfiles, multirow, wrapfig}
\usepackage[font=scriptsize]{subcaption}
\usepackage[font=scriptsize, justification=raggedright]{caption}
\usepackage{subcaption}
\usepackage[toc,page]{appendix}
\usepackage[colorlinks, allcolors=black]{hyperref}
\usepackage[capitalize, nameinlink]{cleveref}
\usepackage{xcolor, tabularx}
\setlist[itemize]{leftmargin=10pt}
\usepackage[utf8]{inputenc}
\usepackage{float}
\usepackage{tikz}
\usetikzlibrary{quantikz}
\usepackage{xcolor}
\hypersetup{
  colorlinks   = true, 
  urlcolor     = black, 
  linkcolor    = teal, 
  citecolor   = teal 
}

\mathcode`l="8000
\begingroup
\makeatletter
\lccode`\~=`\l
\DeclareMathSymbol{\lsb@l}{\mathalpha}{letters}{`l}
\lowercase{\gdef~{\ifnum\the\mathgroup=\m@ne \ell \else \lsb@l \fi}}%
\endgroup

\renewcommand{\epsilon}{\varepsilon}

\newtheorem{lem}{Lemma}

\newtheorem*{thm*}{Theorem}

\renewcommand{\vec}[1]{\overrightarrow{#1}}

\makeatletter
\newcounter{savesection}
\newcounter{apdxsection}
\renewcommand\appendix{\par
  \setcounter{savesection}{\value{section}}%
  \setcounter{section}{\value{apdxsection}}%
  \setcounter{subsection}{0}%
  \gdef\thesection{\@Alph\c@section}}
\newcommand\unappendix{\par
  \setcounter{apdxsection}{\value{section}}%
  \setcounter{section}{\value{savesection}}%
  \setcounter{subsection}{0}%
  \gdef\thesection{\@arabic\c@section}}
\makeatother

\begin{abstract}
     We explore the power of reservoir computing with a single oscillator in learning time series using quantum and classical models. We demonstrate that this scheme learns the Mackey--Glass (MG) chaotic time series, a solution to a delay differential equation. Our results suggest that the quantum nonlinear model is more effective in terms of learning performance compared to a classical non-linear oscillator. We develop approaches for measuring the quantumness of the reservoir during the process, proving that Lee-Jeong's measure of macroscopicity is a non-classicality measure. We note that the evaluation of the Lee-Jeong measure is computationally more efficient than the Wigner negativity. Exploring the relationship between quantumness and performance by examining a broad range of initial states and varying hyperparameters, we observe that quantumness in some cases improves the learning performance. However, our investigation reveals that an indiscriminate increase in quantumness does not consistently lead to improved outcomes, necessitating caution in its application. We discuss this phenomenon and attempt to identify conditions under which a high quantumness results in improved performance.
\end{abstract}

\begin{document}

\title{Quantumness and Learning Performance\\in Reservoir Computing with a Single Oscillator}

\author{Arsalan~Motamedi}
\email{arsalan.motamedi@uwaterloo.ca}
\affiliation{Institute for Quantum Computing, Department of Physics \& Astronomy University of Waterloo, Waterloo, ON, N2L 3G1, Canada}

\author{Hadi~Zadeh-Haghighi}
\email{hadi.zadehhaghighi@ucalgary.ca}
\affiliation{Department of Physics and Astronomy, Institute for Quantum Science and Technology, Quantum Alberta, and Hotchkiss Brain Institute, University of Calgary, Calgary, AB T2N 1N4, Canada}

\author{Christoph Simon}
\email{csimo@ucalgary.ca}
\affiliation{Department of Physics and Astronomy, Institute for Quantum Science and Technology, Quantum Alberta, and Hotchkiss Brain Institute, University of Calgary, Calgary, AB T2N 1N4, Canada}

\date{\today}

\maketitle

\section{Introduction}
The theory of quantum information processing has been thriving over the past few decades, offering various advantages, including efficient algorithms for breaking Rivest–Shamir–Adleman (RSA) encryption, exponential query complexity speed-ups, improvement of sensors and advances in metrology, and the introduction of secure communication protocols \cite{shor1999polynomial, MacQuarrie_2020,Harrow_2009, nielsen2002quantum, bennett2020quantum, degen2017quantum, simon2017towards, rivest1983cryptographic}. Nevertheless, the challenge of error correction and fault-tolerant quantum computing is still the biggest obstacle to the realization of a quantum computer. Despite threshold theorems giving the hope of fault-tolerant computation on quantum hardware \cite{aharonov1997fault, knill1998resilient, kitaev2003fault, shor1996fault}, a successful realization of such methods is only recently accomplished on intermediate-size quantum computers \cite{acharya2022suppressing}, and the implementation of a large scale quantum computer is yet to be achieved. Moreover, today's quantum hardware contain only a few tens of qubits. Hence we are in the noisy intermediate-scale quantum (NISQ) era, and it is of interest to know what tasks could be performed by such limited noisy devices that are hard to do with classical computers \cite{temme2017error, bharti2022noisy, kandala2019error, preskill2018quantum}.

In the past few years, and on the classical computing side, neuromorphic (brain-inspired) computing has shown promising results \cite{farquhar2006field, hopfield1982neural, schmidhuber2015deep, goodfellow2020generative}, most notably the celebrated artificial neural networks used in machine learning. Neuromorphic computing uses a network of neurons to access a vast class of parametrized non-linear functions. Despite being very successful in accuracy, these models are hard to train due to the need to optimize many parameters. Another obstacle in the training of such models is the vanishing gradient problem \cite{pascanu2013difficulty, basodi2020gradient}.

A subfield of brain-inspired computing, derived from recurrent neural networks, is reservoir computing, where the learning is to be performed only at the readout. Notably, this simplification - optimizing over a smaller set of parameters - allows for circumventing the problem of barren plateaus encountered in the training of recurrent neural networks. Despite such simplifications, reservoir computing still shows remarkable performance \cite{maass2002real,jaeger2004harnessing,tanaka2019recent, rohm2018multiplexed, nature1, nakajima2018reservoir}. Reservoir computing methods are often applied to temporal categorization, regression, and also time series prediction \cite{schrauwen2007overview, mammedov2022weather}. Notably, there have been successful efforts on the physical implementation of (classical) reservoir computing \cite{tanaka2019recent, kan2022physical, nakajima2018reservoir}. 

More recently, the usefulness of quantum computing in the field of machine learning has been studied \cite{biamonte2017quantum, QML, schuld2015introduction}. In addition to that, there are novel attempts to introduce an appropriate quantum counterpart for classical neuromorphic (in particular reservoir) computing. There have been different reservoir models considered, which could mostly be categorized as spin-based or oscillator-based \cite{fujii2021quantum,PhysRevResearch.3.013077, luchnikov2019simulation, martinez2020information, nokkala2021gaussian} (corresponding to finite and infinite dimensional Hilbert spaces). 

On the quantum reservoir computing front, there have been efforts such as \cite{PhysRevResearch.3.013077}, where one single Kerr oscillator is exploited for fundamental signal processing tasks. The approach used in \cite{nokkala2021gaussian} for quantum reservoir computing introduces non-linearity through the encoding of input signals in Gaussian states. Their approach has been proven to be universal, meaning that it can accurately approximate fading memory functions with arbitrary precision. \cite{pfeffer2022quantum} predicts time series using a spin-based model. \cite{ghosh2021quantum} exploits a network of interacting quantum reservoirs for tasks like quantum state preparation and tomography. \cite{vintskevich2022computing} proposes heuristic approaches for optimized coupling of two quantum reservoirs. An analysis of the effect of quantumness is performed in \cite{PhysRevResearch.3.013077}, where the authors consider dimensionality as a quantum resource. The effects of quantumness have been studied more concretely in \cite{gotting2023exploring}, where they consider an Ising model as their reservoir and show that the dimension of the phase space used in the computation is linked to the system's entanglement. Also, \cite{pfeffer2022quantum} demonstrates that quantum entanglement might enhance reservoir computing. Specifically, they show that a quantum model with a few strongly entangled qubits could perform as well as a classical reservoir with thousands of perceptions, and moreover, performance declines when the reservoir is decomposed into separable subsets of qubits. 

In this work, we explore how well a single quantum non-linear oscillator performs time series predictions. In particular, we are focused on the prediction of the Mackey-Glass (MG) time series \cite{mackey1977oscillation}, which is often used as a benchmark task in reservoir computing. We then investigate the role of quantumness in the quantum learning model. We use Lee-Jeong measure \cite{lee2011quantification} as a measure of quantumness. This measure was originally introduced for macroscopicity, but here we demonstrate that it is a non-classicality measure as well. We highlight that Lee-Jeong measure can be computed efficiently, as opposed to Wigner negativity. Using our approaches, we observe that quantumness can enhance performance, measured in terms of test error. Nevertheless, more quantumness does not always lead to improved performance; caution is necessary. In particular, our results demonstrate that there exists a pattern in the interplay of hyperparameters (with a proper choice of initial state), indicating optimal performance within specific parameter ranges, for which we provide an intuitive reason. Moreover, it is the case that certain states consistently yield poor learning performance across a broad spectrum of hyperparameters.

The paper is organized as follows. In \cref{sec:rc} we introduce the reservoir computing method used in this work. \cref{sec:pts} shows the performance of the method. \cref{sec:q} analyzes the effect of quantumness measures on performance. Finally, \cref{sec:discussion} provides a discussion of the findings.

\begin{figure}[t]
    \centering
    \includegraphics[width = 0.4\textwidth]{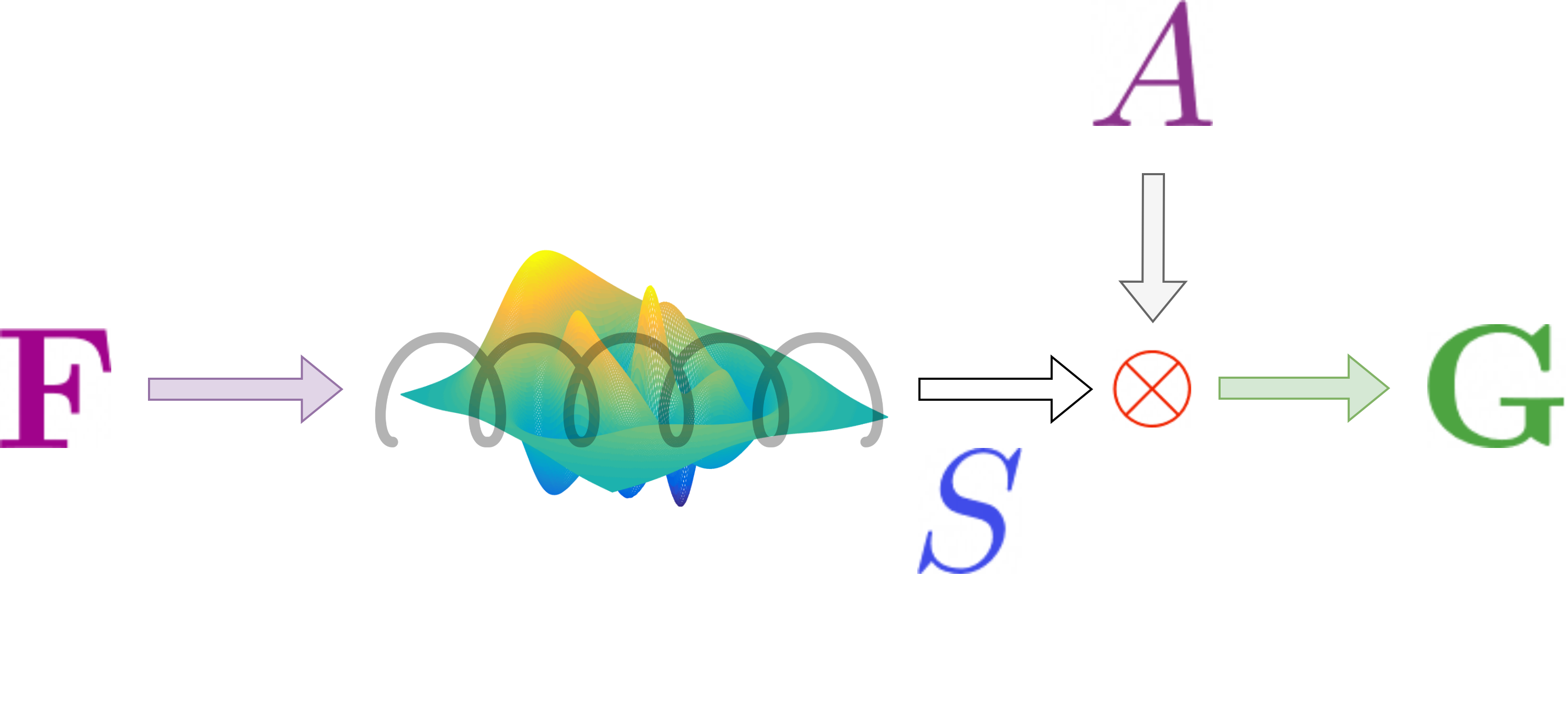}
    \caption{A schematic representation of the computation model, either classical or quantum. In the learning process, we find the proper $A$ that is to predict the sample set $G$, based on the outputs of the reservoir. The dynamics of the reservoir is controlled via sample set $F$.}
    \label{fig:Schem}
\end{figure}

\begin{figure*}[!tbp]
  \centering
  \subfloat[The oscillator is trained to reproduce the Mackey-Glass time series, and then predict, given initial values, which are the data points before the dashed line.]{\includegraphics[width=0.4\textwidth]{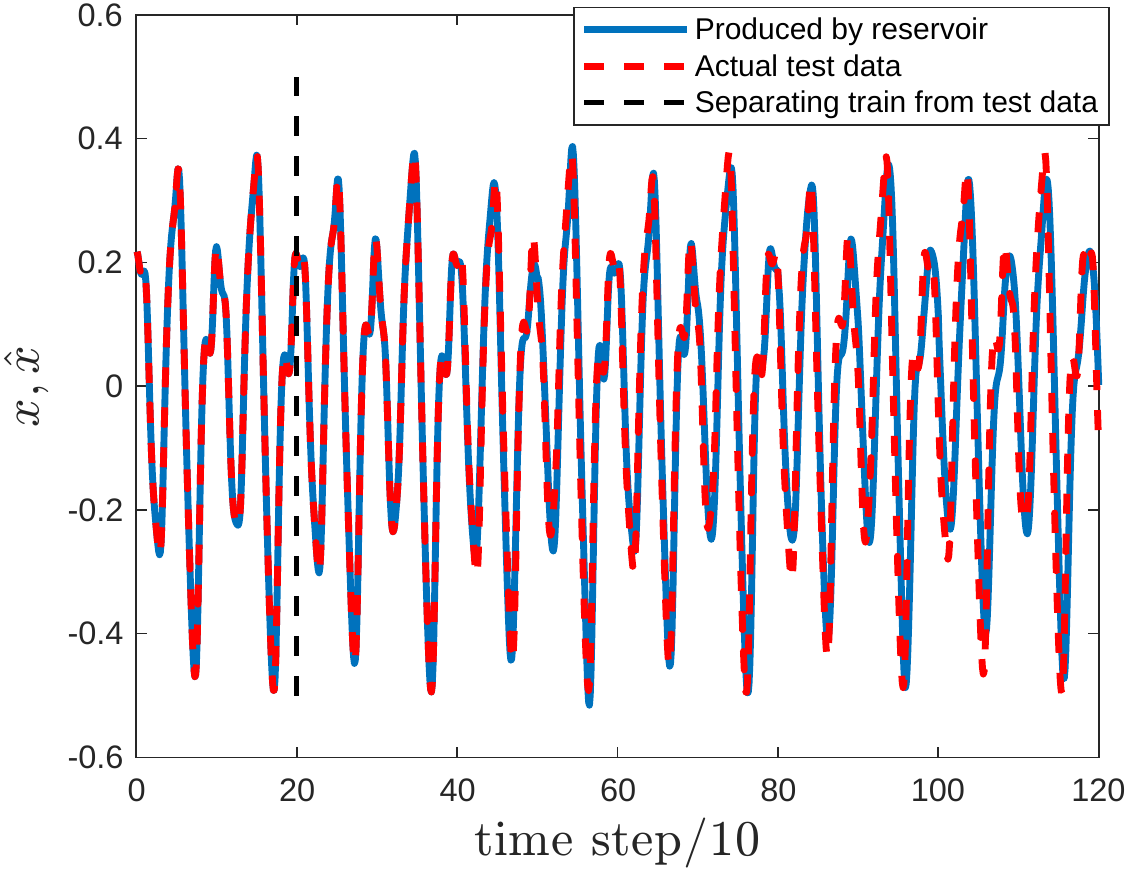}\label{fig:MG1}}
  \hspace{0.2 cm}
  \subfloat[Comparing the delayed embedding of the actual Mackey-Glass series (left) with that of the reservoir's output (right). See \cref{sec:pts} for definitions.]{\includegraphics[width=0.4\textwidth]{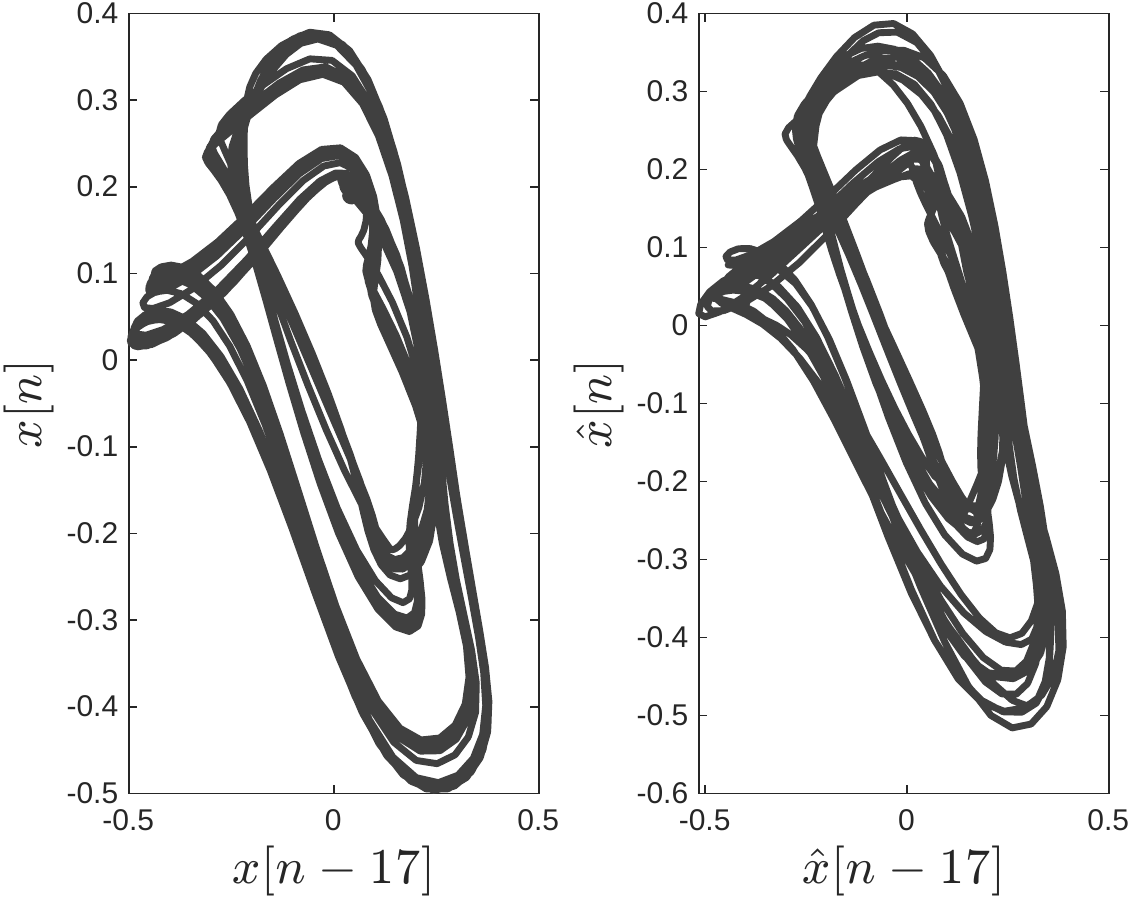}\label{fig:MG2}}
\caption{Performance of the trained quantum model. Here $n$ denotes the time step or equivalently the index of the sampled point from the function, $x$ and $\hat x$ refer to the actual and the predicted series, respectively.}
\label{fig:1}
\end{figure*}

\section{Reservoir Computing}\label{sec:rc}
In this work, we use the approach introduced by Govia et al. \cite{PhysRevResearch.3.013077} to feed the input signal to the reservoir by manipulating the Hamiltonian. We describe the state of our quantum and classical systems using $\hat\rho$ and $a$, respectively. Let us consider a time interval of length $\Delta t$ that has been discretized with $N$ equidistant points $t_1< \cdots< t_N$. We can expand this set by considering $M$ future values $t_{N+1}< \cdots< t_{N+M}$, which are also equidistantly distributed. It is worth noting that the interval $t_N - t_1$ is equal to $\Delta t$. Our objective is to estimate a set of future values of the function $f$, which is denoted by $G = \left(f(t_{i})\right)_{i=N+1}^M$, given the recent observations $F=\left(f(t_i)\right)_{i=1}^N$. To this end, we evolve our system so that $\hat\rho(t_j)$ (respectively, $a(t_j)$) depends on $f(t_1), \cdots, f(t_j)$. We then obtain observations $s(t_i) = \langle\hat{O} \hat\rho(t_i)\rangle$ for an observable $\hat O$ (respectively, $s(t_i) = \left(h \circ a\right){t_i}$ for a function $h$). Finally, we perform a linear regression on $\left( s(t_i)\right)_{i=1}^N$ to predict $G$. In what follows we elaborate on the system's evolution in both classical and quantum cases.

For the classical reservoir, we consider the following evolution
    \begin{align}\label{eq:classicEv}
        \dot{a} = -iK(1+2\, |a|^2)a - \frac{\kappa}{2} a - i\alpha f(t)
    \end{align}
    with $K$, $\kappa$, and $\alpha$ being the reservoir's natural frequency, dissipation rate, and the amplifier of the input signal, respectively. We let $s(t) = \tanh\left(\Re{a(t)}\right)$. The quantum counterpart of the evolution described by \cref{eq:classicEv} is the following Markovian dynamics 
 \cite{PhysRevResearch.3.013077,gardiner2004quantum}
\begin{equation}
\begin{split}
    \frac{\mathrm d}{\mathrm dt}\hat\rho(t) &= -i [\hat{H}(t), \hat\rho(t)] + \kappa\, \mathcal{D}_a(\hat\rho)\\
    \text{where }\, \hat{H}(t) &= K\, \hat{N}^2 + \alpha \, f(t)\, \hat{X}, \\ \text{and }\, \mathcal{D}_a(\hat\rho) &= \hat{a} \hat\rho \hat{a}^\dagger - \{\hat{N}, \hat\rho\}/2. \label{eq:ev-2}
\end{split}
\end{equation}
Here $\hat a$ refers to the annihilation operator and is related to $\hat X$ and $\hat N$ through $\hat X = \hat a+\hat a^\dagger$ and $\hat N = \hat a^\dagger \hat a$.
Note that we use the properly scaled parameters, such that the uncertainty principle becomes $\Delta x \, \Delta p \geq 1/2$ (we have employed the notation $\Delta A = \langle (\hat A-\langle \hat A\rangle)^2\rangle$ for an observable $\hat A$). The parameters $(\alpha, \kappa, K)$ are the same as above (in \eqref{eq:classicEv}). The operators $\hat X$, $\hat a$, and $\hat N$ represent the position, annihilation, and the number operator respectively. We let $s(t) = \tr ( \hat\rho(t) \,\tanh \hat X)$. Utilizing the non-linear quantum evolution \eqref{eq:ev-2} as our quantum reservoir, we perform the learning of the MG time series. Throughout, in time series prediction, both $f$ and $g$ refer to the same series, that is generally denoted by $x$.

Let us now provide details on the process of linear regression. Our objective is to find the predictor $A$ that satisfies the relationship
$
G \approx A s
$
(note that we think of $G$ and $s$ as column vectors). To this end, we conduct the experiment $T$ times and collect the resulting column vectors as ${ \vec s_1, \vec s_2, \cdots, \vec s_T}$. We then define a matrix $\mathbf{S}$ as the concatenation of these column vectors
\begin{align}
    \mathbf{S} := \begin{pmatrix}
     s_1 |\,  s_2 |\, \cdots | s_T
    \end{pmatrix}
\end{align}
Similarly, we define the matrix $\mathbf{G}$ as
\begin{align}
    \mathbf G := \begin{pmatrix}
     G_1 |  G_2|\, \cdots \, |  G_T
    \end{pmatrix}.
\end{align}
Finally, we choose $A$ by applying Tikhonov regularization \cite{shalev2014understanding}, which results in the following choice of $A$
\begin{align}\label{eq:W}
A =  \mathbf G \mathbf{S}^T (\mathbf{S}\mathbf{S}^T + {\eta} \mathbb{I})^{-1}
\end{align}
with $\eta$ and $\mathbb I$ being a regularization parameter, and the identity matrix, respectively. One should note that $\mathbf G$ and $\mathbf S$ written in \eqref{eq:W} correspond to $T$ training samples that we take. The matrix $A$ evaluated above is then used for the prediction of the test data. \cref{fig:Schem} provides a schematic representation of the reservoir training.

Overall, to predict time series, the reservoir is initially trained to determine $A$ by using equation \eqref{eq:W}. After training, a set of initial values outside of the training data is inputted into the oscillator. The oscillator uses $A$ to predict future values, which are then used as initial values for further predictions.

\section{Results}\label{sec:res}
This section investigates the performance of a single Kerr non-linear oscillator trained on the MG chaotic time series, as well as the effect of quantumness on the performance of the reservoir. Specifically, in \cref{sec:pts}, we discuss how well the non-linear oscillator learns chaotic time series, and in \cref{sec:q}, we examine the impact quantumness on the performance. Lastly, we outline further investigations, including the effects of noise.

To simulate quantum dynamics in the Fock space, we truncate every operator in the number basis, making them $d_{\text{t}}$-dimensional. Notably, the simulation results in this work use a dimension $d_{\text{t}}\geq 20$, which is sufficiently large as most of the states considered have a significant overlap with the subspace spanned by the number states $\ket n$ for $n\leq10$ (For instance, we use the coherent state $\ket\alpha$ with $\alpha=1+i$, the overlap of which with the first $20$ Fock states is larger than $1-6.5\times 10^{-15}$).

\subsection{Learning Time Series}\label{sec:pts}
In what follows we report the results obtained by training our single non-linear oscillator.
Here, we consider the prediction of the chaotic MG series, which is formally defined as the solution to the following delay differential equation
\begin{align}\label{eq:MG-evolutino}
\dot{x}(t) = \beta \frac{x(t-\tau)}{1+x(t-\tau)^m} - \gamma x(t).
\end{align}
We use the parameters $\beta = 0.2$, $\gamma = 0.1$, $m=10$, and $\tau = 17$ throughout. In \cref{app:changing-tau} of our Supplementary Material, we discuss that $\tau$ controls the complexity of the series, and we repeat the training for different values of $\tau$. We demonstrate that learning a series with a larger $\tau$ is still feasible if we increase the input length $N$ (as defined in \cref{sec:rc}). The performance of the trained reservoir on the test data is presented in \cref{fig:MG1}. \cref{fig:MG2} shows the delayed embedding of the predicted MG, which is compared to the actual diagram. One can readily observe that this model is successful in learning MG. We emphasize that we are using the units in which $h=2\pi$. We then integrate the evolution of the reservoir with time step $\Delta t=0.1$ in that unit (this is identical to $\Delta t$ introduced in \cref{sec:rc}). Moreover, using \eqref{eq:MG-evolutino}, we generated the Mackey--Glass series, and sampled it at the integer points in time $t$. This provides us with a discrete set of data that is then fed to the reservoir.

\cref{table:learning-params} summarizes the parameters employed in the training of the model, by which we generated \cref{fig:MG1} and \cref{fig:MG2}.

\begin{table}[h]
  \centering
  \caption{Learning parameters and values that are used in obtaining \cref{fig:MG1} and \cref{fig:MG2}. Please refer to \cref{sec:rc} for a detailed explanation of these parameters. The units of the parameters is consistent with the formulation of \eqref{eq:ev-2}.}
  \label{tab:learning-parameters}
  \begin{tabular}{l l l}
    \toprule
    Parameter & Description & Value \\
    \midrule
    $K$ & non-linearity strength & $0.05$ \\
    $\kappa$ & dissipation rate & $0.1$ \\
    $\alpha$ & input coefficient & $1.2$ \\
    $N$ & input length & $200$ \\
    $M$ & output length & $100$ \\
    $\Delta t$ & time step & $0.1$ \\
    $\eta$ & regularization parameter & $0.01$ \\
    $\ket{\psi_{\mathrm{init}}}$ & initial state of reservoir & $\ket 6$\\
    $T$ & number of training rounds & 248\\
    \bottomrule
  \end{tabular}
  \label{table:learning-params}
\end{table}

\subsection{Quantumness}\label{sec:q}
In this section, we introduce our quantumness measure and study the effect of quantumness on that basis on the accuracy of the learning model.
\begin{figure}
    \centering
    \includegraphics[width=0.5\textwidth]{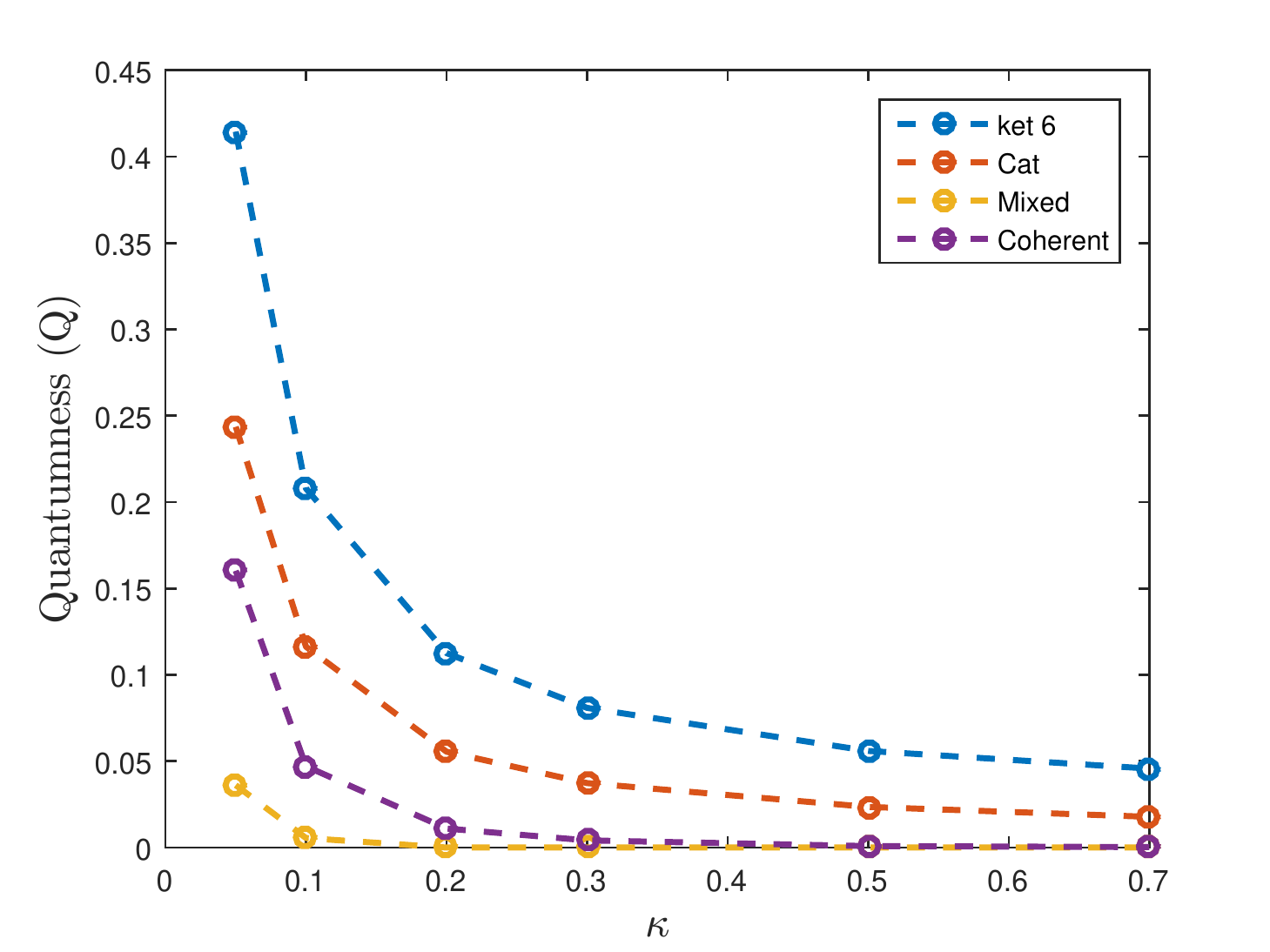}
    \caption{
    Average quantumness ($Q$) during the evolution is shown to be decreasing as $\kappa$ (the photon loss rate) increases. The states used as initial states are the `Coherent' state $\ket\alpha$, the mixed state (labeled as `mix') being proportional to $\ket{\alpha}\bra{\alpha} + \ket{-\alpha}\bra{-\alpha}$, the corresponding `cat' state being proportional to $\ket{\alpha}+\ket{-\alpha}$, and the $\ket{6}$ (`ket 6'). The parameters are $(\alpha, \lambda, K) = (1.2, 0.1, 0.05)$, where $\lambda$ is the pumping parameters, used to keep all the quantumness values in a similar range. We use $\alpha = 1+i$ for the coherent, the mixed, and the cat states. Please refer to \cref{SM:noise} of the Supplementary Material for the definition of the noise model corresponding to the parameter $\lambda$.}
    \label{fig:Wigs}
\end{figure}

Our main goal here is to determine if there is a relation between the quantumness of the system and the accuracy of the learning process. We aim to explore whether quantumness can serve as a valuable resource for reservoir computing. To this end, we need to quantify the quantumness of a state in Fock space. We point out that there has been extensive research done on the quantification of quantumness \cite{ groisman2007quantumness, ollivier2001quantum, takahashi1986wigner}. 
Furthermore, there has been a line of research in the study of the macroscopicity of quantum states and their effective size \cite{frowis2018macroscopic, nimmrichter2013macroscopicity, leggett1985quantum, leggett2016note}. One such measure in the Fock space is defined by
Lee and Jeong \cite{lee2011quantification} as follows:

\begin{equation}
\begin{split}
    I(\hat\rho) := \pi  &\int\bigg( \big(\partial_xW(x,p)\big)^2\\ &+\big(\partial_pW(x,p)\big)^2 - 2W(x,p)^2 \bigg) \, {\mathrm dx\,\mathrm dp}
\end{split}
\end{equation}
where $x,p$ are dimensionless space and momentum components (in such scale, the uncertainty principle becomes $\Delta x\, \Delta p \geq 1/2$). Here $W(x,p)$ refers to the Wigner function
\begin{align}\label{eq:def-wig}
W(x, p) = \frac{1}{\pi} \int_{-\infty}^{\infty} \langle x - \frac{y}{2} | \hat{\rho} | x + \frac{y}{2} \rangle e^{ipy} \, \mathrm dy.
\end{align}
The above formulation of $I(\hat\rho)$ can also be found in \cite{I}. The following identities are pointed out in \cite{lee2011quantification}:

\begin{itemize}
    \item $I(\ket{\alpha} \bra{\alpha}) = 0$, for any coherent state $\ket\alpha$.
    \item $\forall n \in \mathbb{N}: I(\sum_{i=0}^{n-1} \frac{1}{n}\, \ket{i}\bra{i}) = 0$, where $\ket{i}$ are the Fock states (i.e., the eigen-vectors of the number operator).
\end{itemize}

It is worth mentioning that this measure can obtain negative values \cite{I}. Intuitively, one could think of $I(\hat\rho)$ as the fineness of the Wigner function associated with $\hat\rho$. The aforementioned results may suggest that the positiveness of $I$ indicates non-classicality, as the coherent state and a diagonal density matrix in the Fock basis are considered as classical states. In the following theorem, we prove that if $I(\hat\rho) >0$, then $\hat\rho$ cannot be written as a mixture of coherent states, hence being non-classical.

In what follows, we prove that if $\hat\rho$ is a mixture of coherent states, then $I(\hat\rho)\leq 0$. For any coherent state $\ket{\alpha}$, one has 
\[
W_{\alpha} = \frac{1}{\pi}\, e^{- \big[ (x-\text{Re}(\alpha))^2 + (p-\text{Im}(\alpha))^2 \big]}.
\]
Let us consider a set of $K$ coherent states, namely $\{\hat\rho_{i} = \ket{\alpha_i}\bra{\alpha_i}: i=1,2,\cdots,K  \}$, and define $x_i : = \text{Re}(\alpha_i)$, $p_i := \text{Im}(\alpha_i)$. Also let $(q_i)_{i\in [K]}$ be a probability distribution over $K$ objects. We can then consider the mixture of coherent states as
\begin{align*}
    \hat\rho = \sum_{i=1}^K q_i\, \ket{\alpha_i}\bra{\alpha_i}
\end{align*}
since the Wigner function is linear with respect to the density matrix, one has
$
W_{\hat\rho}(x,p) = \sum_{i=1}^K q_i\, W_{\alpha_i}(x,p).
$
Hence, we get
\begin{align*}
    &\frac{1}{\pi}\, I(\hat\rho) \\
    &= \int \left(\sum_{i\in[K]} \, q_i\,  \partial_xW_i\right)^2
      + \left(\sum_{i\in[K]} \, q_i\, \partial_pW_i\right)^2 - 2 \left(\sum_{i\in[K]} \, q_i\, W_i\right)^2  {\mathrm dx\mathrm dp}\\
    &= \sum_{i,j\in[k]} \, q_i\, q_j\, \int\bigg( \partial_{x} W_i\, \partial_{x} W_j+ \partial_{p} W_i\, \partial_{p} W_j - 2W_i\, W_j \bigg) {\mathrm dx\mathrm dp}
\end{align*}
note that the terms in the summation above with $i=j$ could be rewritten as $q_i^2\, I(\hat\rho_i) = 0$ since any cohrent state $\hat\rho_i$ has zero quantumness i.e., $I(\hat\rho_i)=0$. Furthermore, our explanation below guarantees that for any choice of $i,j$ the expression in the parenthesis is non-positive, and hence, the proof is complete.

All that is left, is to prove that for any two coherent states, say $\ket{\alpha_0}$ and $\ket{\alpha_1}$, both of the following inequalities hold
\begin{equation}
\begin{split}\label{eq:inequality}
\int \bigg(\partial_{x} W_0\, \partial_{x} W_1 - W_0\, W_1\bigg){\mathrm dx\mathrm dp} &\leq 0,\\
\int  \bigg(\partial_{p} W_0\, \partial_{p} W_1 - W_0\, W_1\bigg){\mathrm dx\mathrm dp} &\leq 0
\end{split}
\end{equation}
We start by writing
\begin{align*}
    W_0 = \frac{1}{\pi}\, e^{- \big[ (x-x_0)^2 + (p-p_0)^2  \big]}
\end{align*}
hence
\begin{equation*}
\begin{split}
    \partial_x W_0 &= -2 \frac{(x-x_0)}{\pi}\, e^{- \big[ (x-x_0)^2 + (p-p_0)^2  \big]},\\ \partial_p W_0 &= -2 \frac{(p-p_0)}{\pi}\, e^{- \big[ (x-x_0)^2 + (p-p_0)^2  \big]},
\end{split}
\end{equation*}
and similar expressions for $W_1$ and its derivatives. Let us now prove the first inequality. Define
\begin{align*}
    \mathcal A := \int \bigg(\partial_{x} W_0\, \partial_{x} W_1 - W_0\, W_1\bigg) \, {\mathrm dx \, \mathrm dp}
\end{align*}
then, by direct substitution one gets
\begin{align*}
    \mathcal A = \frac{1}{\pi^2}\, &\int \big( 4(x-x_0)(x-x_1) - 1 \big) \notag \\
    & \times e^{- \big[ (x-x_0)^2 + (x-x_1)^2 + (p-p_0)^2 + (p-p_1)^2 \big] } {\mathrm dx\mathrm dp},
\end{align*}
we may now use the elementary identities
$
    (x-x_0)(x-x_1) =\left(x - \frac{x_0+x_1}{2}\right)^2 - \left(\frac{x_0 - x_1}{2} \right)^2$ and $
    (x-x_0)^2 + (x-x_1)^2 =\left(x - \frac{x_0+x_1}{2}\right)^2 + \left(\frac{x_0 - x_1}{2} \right)^2
$
and further, letting $\Delta x := x_0 - x_1$, $\Delta p = p_0-p_1$, and $\overline{x} := \frac{x_0 + x_1}{2}$, and $\overline{p}= \frac{p_0+p_1}{2}$ to conclude
\begin{align*}
\mathcal A &= \frac{e^{-\frac{1}{2} ( \Delta x^2 + \Delta p^2)  }}{\pi^2} \int  \big[ 4\big(x - \overline{x}\big)^2 - (\Delta x)^2 - 1 \big] e^{ -2(x - \overline{x})^2 -2(p - \overline{p})^2 }\\
&= -\frac{e^{-\frac{1}{2} ( \Delta x^2 + \Delta p^2)}}{2\pi}\, (\Delta x)^2 \leq 0,
\end{align*}
where the last equality follows from elementary Gaussian integrals. A similar argument gives the second inequality of \eqref{eq:inequality}.

\begin{figure}[h]
    \centering
    \subfloat[Two snapshots of Wigner functions of the reservoir's state, starting from different states. The plots correspond to the initial state being the mixed state (i.e. proportional to $\ket{\alpha}\bra{\alpha}+\ket{-\alpha}\bra{-\alpha}$), cat state (i.e. proportional to $\ket{\alpha}+\ket{-\alpha}$), a coherent state $\ket\alpha$, and the number state $\ket 6$. Here, $X$ refers to position and $P$ refers to momentum (see \eqref{eq:def-wig} for the definition).]{\includegraphics[width = 0.4\textwidth]{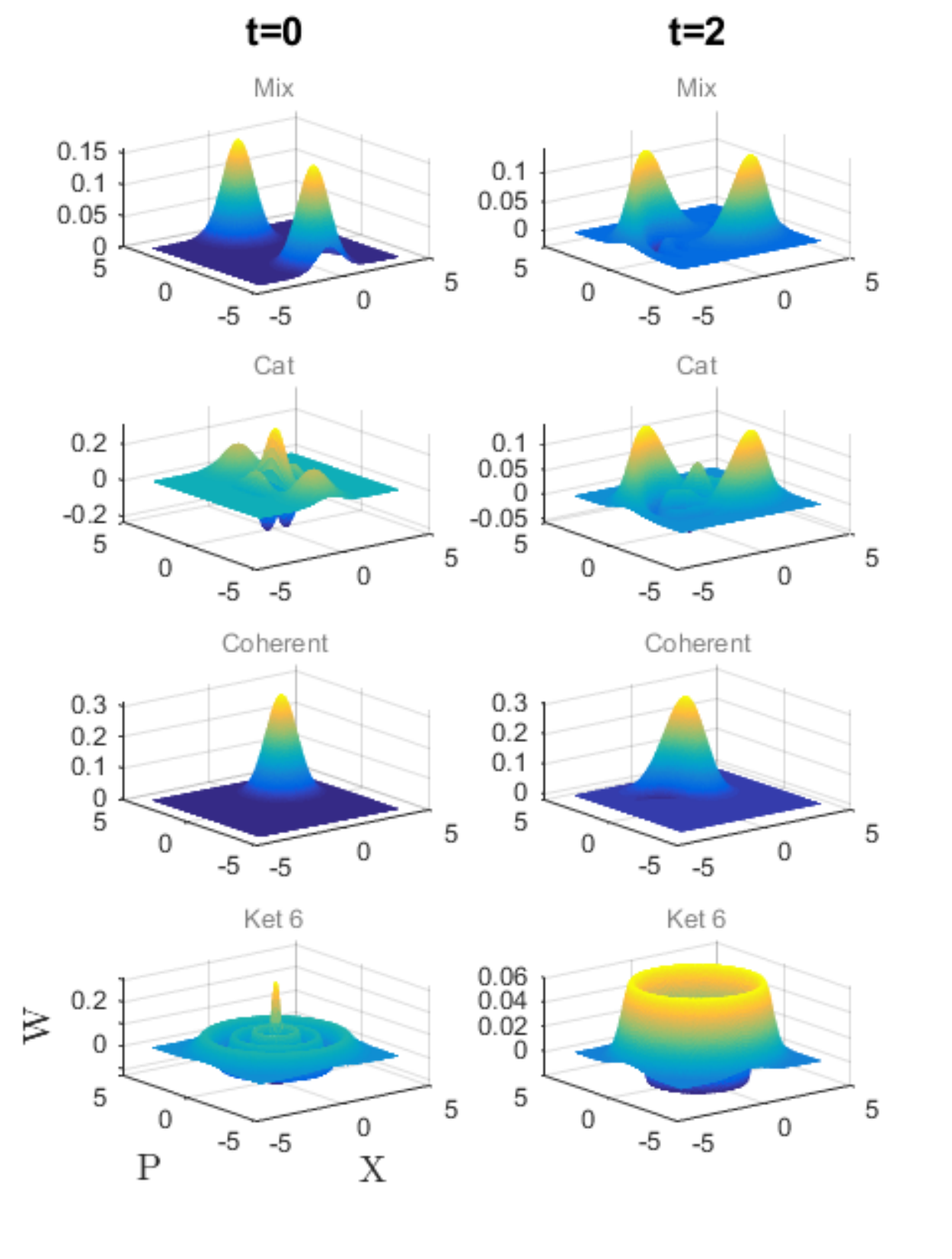}}
    \hfill
    \subfloat[Quantumness (Q) during the evolution. The label 'mix' corresponds to the case where the initial state is $\ket{\alpha}\bra{\alpha} + \ket{-\alpha}\bra{-\alpha}$, the label 'cat' corresponds to $\ket{\alpha} + \ket{-\alpha}$, and the label 'ket6' corresponds to $\ket 6$ (Note that the correspondences are up to a normalization factor). The values are normalized by the absolute maximum in each diagram.]{\includegraphics[width = 0.4\textwidth]{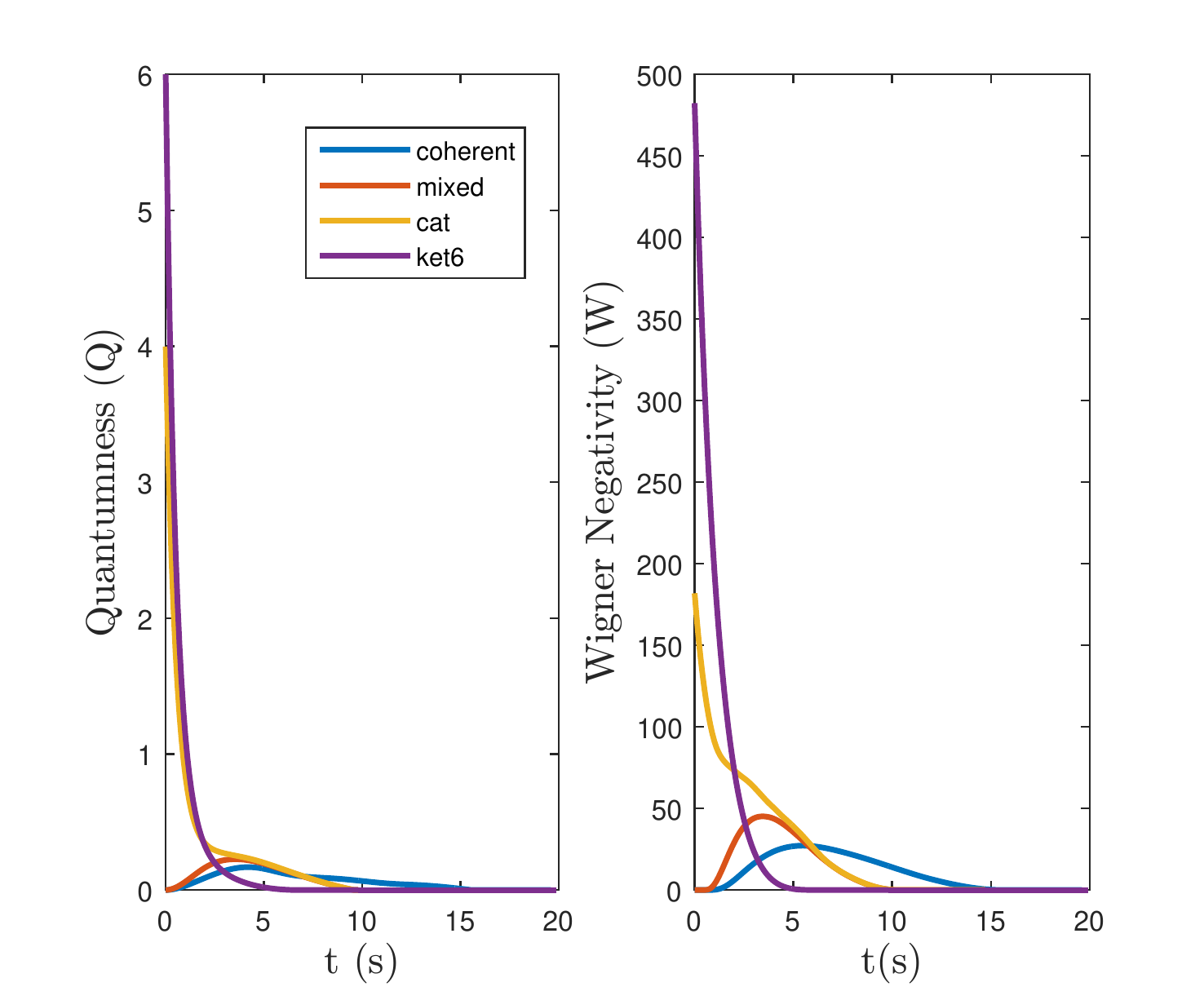}}
    \caption{Quantumness and Wigner plots of the reservoir's state evolution. The oscillator parameters are $(\alpha, \kappa, K) = (1.2, 0.1, 0.05)$.}
    \label{fig:QuantKappa}
\end{figure}

One should also note that $I$ is computable in a much shorter time since it can be reformulated as (\cite{lee2011quantification})
\begin{align}
I(\hat\rho) = -\tr\left(\hat\rho \mathcal{D}_a(\hat\rho)\right).
\end{align}
On the other hand, the computation of Wigner negativity with the current algorithms is costly, as it requires the computation of the entire Wigner function.
We hence use the following quantumness measure $Q$
\begin{align}
Q(\hat\rho) = \begin{cases}
I(\hat\rho) \, &\text{if } I(\hat\rho) >0,\\
0 & \text{o.w.}
\end{cases}
\end{align}
We make this choice as we do not want our quantumness measure to obtain negative values. We observe that this measure is consistent with some intuitions regarding the quantumness of reservoir computing, which have been previously used in \cite{PhysRevResearch.3.013077}. In particular, the intuition that by increasing $\kappa$ we should reach a classical limit, which is illustrated in \cref{fig:Wigs}. Finally, we point out the fact that the quantumness of the state changes during evolution. This is indeed observed in \cref{fig:QuantKappa}.

\subsection{Exploring parameter space}\label{sec:ParamSpace}
The set of hyperparameters define the dynamics of an oscillator through \eqref{eq:ev-2}, and therefore, the quantumness exhibited by the model depends on these parameters. As previously discussed (e.g. see \cref{fig:Wigs}), a large value for $\kappa$ yields a classical oscillator, while high quantumness is expected when $\kappa$ is significantly smaller than $K$. It is worth noting that $\alpha$ in \eqref{eq:ev-2} dictates the responsiveness to input data. In this section, we maintain $\alpha$ at a constant value of $1.2$, consistent with the experiments conducted in preceding sections, while exploring the impact of other parameters ($K$ and $\kappa$) on the learning process.

The simulation methodology is structured as follows: We select pairs of $(K, \kappa)$ where $K$ takes values from $\{0.02, 0.05, 0.07, 0.1, 0.12\}$ and $\kappa$ from $\{0.02, 0.03, 0.05, 0.1, 0.2, 0.3\}$. For each pair, the oscillator is initialized at a state from a collection of 35 random states, one at a time. It is important to note that the same set of random states is utilized across all pairs of $(K, \kappa)$. For the initial random states, we fix a dimension $d$ (which determines its support on the Fock basis), then pick a state according to the Haar random measure on $\mathcal H(\mathbb C^d)$ \cite{haar1933massbegriff}.
 To this end, we use \cite[Proposition 7.17]{watrous2018theory}. In particular, we consider a set of $2d$ independent and identically distributed (i.i.d.) standard Gaussian random variables $\zeta_1,\cdots,\zeta_{2d}$ to construct the state

 \begin{align}\label{eq:Gaussian}
 \ket{\psi} = \frac{1}{\sqrt{\sum_{m=0}^{2d-1}\zeta_m^2}}\sum_{n=0}^{d-1} (\zeta_{2n} + i\zeta_{2n+1}) \ket n
 \end{align}
 where $\ket n$ are the Fock basis states. We highlight that due to \cite[Proposition 7.17]{watrous2018theory}, the state $\ket\psi$ is a Haar random state in $\mathcal H(\mathbb C^d)$, meaning that its distribution is invariant under the action of the unitaries acting on $\mathcal{H}(\mathbb C^d)$. Through this method, we generate $5$ random states for each dimension $d$, ranging from $4$ to $10$, resulting in a total of $35$ states.

\begin{figure}[t]
    \centering
    \includegraphics[width=.45\textwidth]{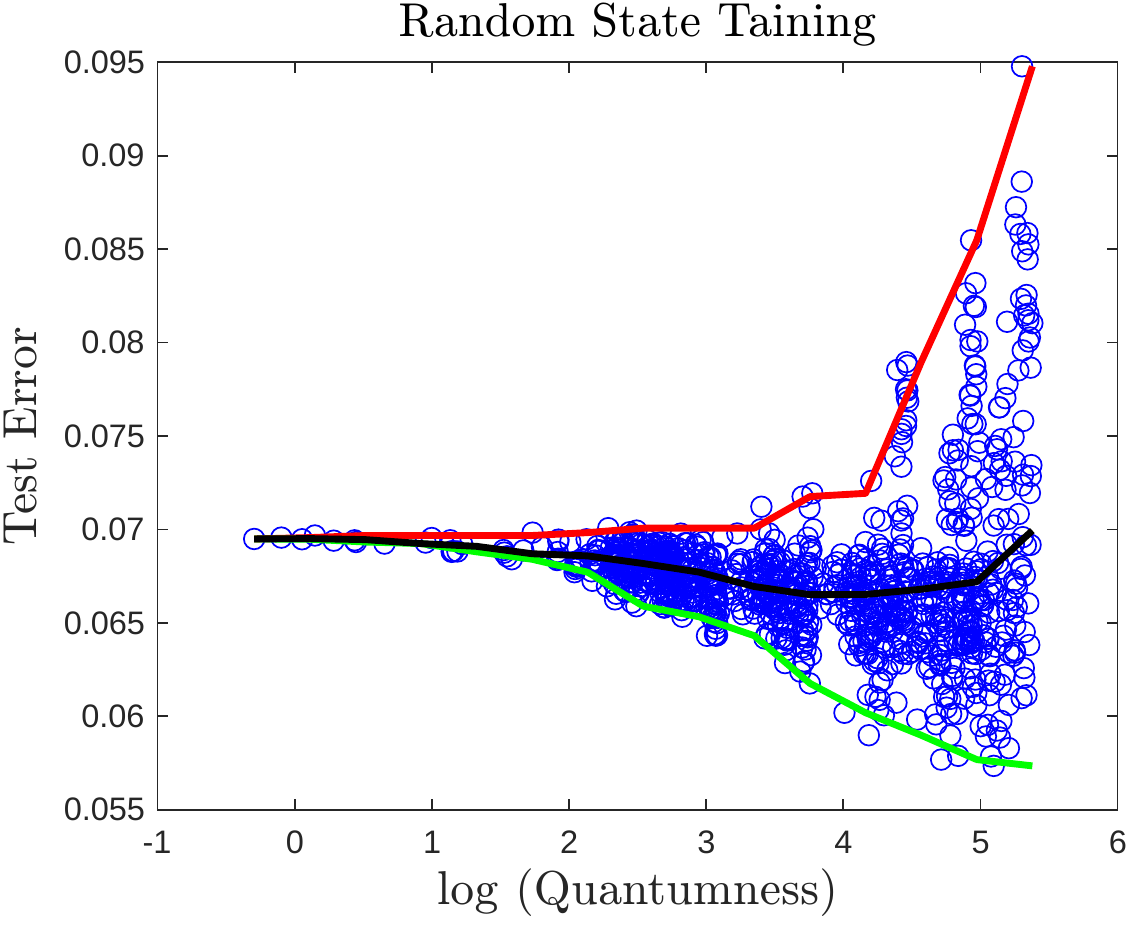}
    \caption{All data points obtained from the task of \cref{sec:ParamSpace}. Each blue point corresponds to the outcome of the training on a particular state with a particular set of reservoir parameters $(K,\kappa)$. The solid red and green curves are piece-wise linear estimators to the worst and best achieved test error at each quantumness value, respectively. The solid black shows the average test error at each quantumness value. We have employed the natural logarithm in the scaling of the horizontal axis.}
    \label{fig:ParamSpace}
\end{figure}

\begin{figure}[h]
    \centering
    \includegraphics[width=.45\textwidth]{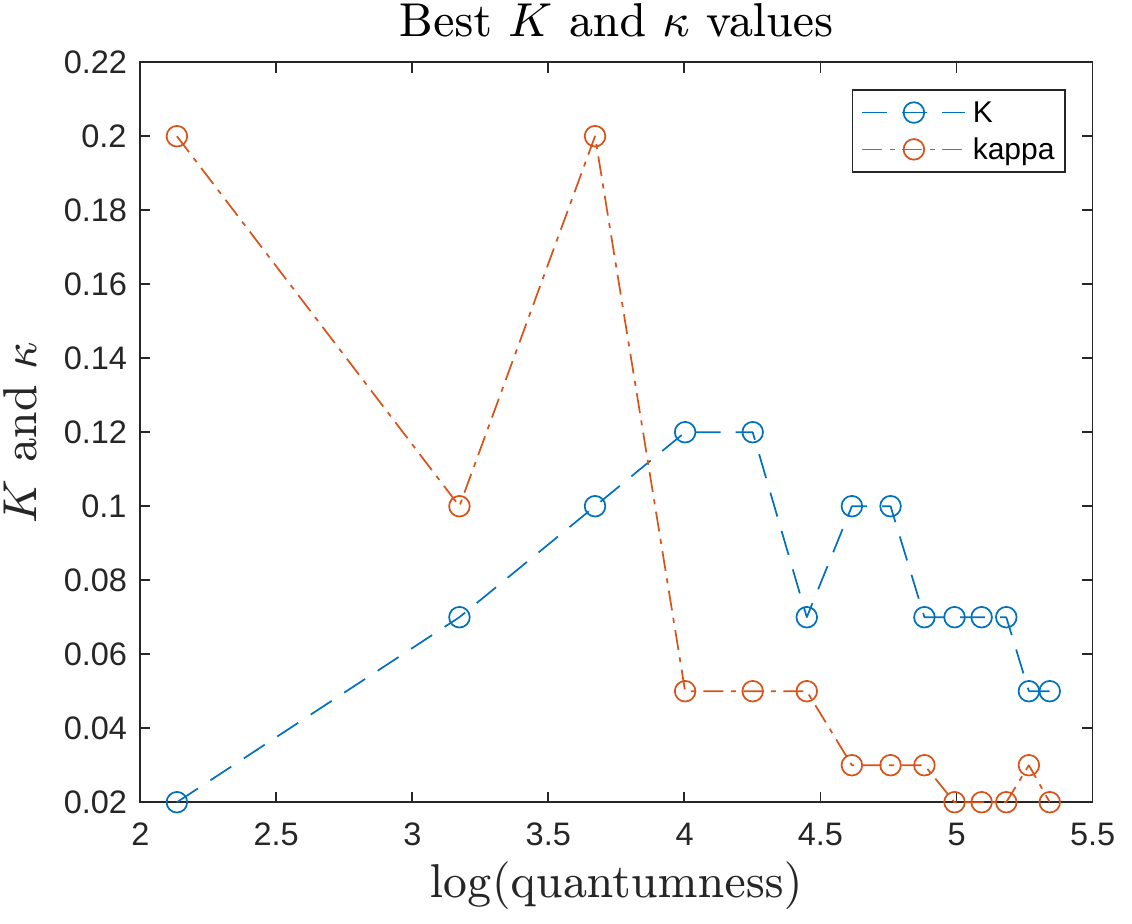}
    \caption{Values of $K\in \{ 0.02, 0.05, 0.07, 0.1, 0.12 \}$ and $\kappa\in \{ 0.02,0.03,0.05,0.1,0.2,0.3 \}$ which correspond to the best performing reservoir, as a function of quantumness. Note that the absolute best reservoir corresponds to $K=0.05, \kappa = 0.02$. The logarithm for the x-axis scaling is the natural logarithm, i.e., base $e$.}
    \label{fig:bestKk}
\end{figure}

The described approach yields the results presented in \cref{fig:ParamSpace}. Furthermore, \cref{tab:full-data} provides comprehensive details regarding this simulation. Our first observation is that quantumness has the potential to improve performance, as measured by test error. However, an increase in quantumness does not guarantee a corresponding improvement in performance. In addition to the quantumness-performance relationship, there are other aspects to consider, such as identifying the optimal parameter regime $(K, \kappa)$ for achieving peak performance. From the minimum error values in \cref{tab:full-data} and also from \cref{fig:bestKk}, the optimal performance occurs when $\kappa \ll K \ll \alpha$. This outcome is intuitively explained by the fact that $\kappa \ll K$ corresponds to high quantumness, which is the regime with best performance. Also $\kappa, K \ll \alpha$ indicates the dominance of dynamics generated by the input data, facilitating the reservoir's learning process. It is evident that if the reservoir's dependency on input data is obfuscated by other dynamics (even in the high quantumness regime), its ability to  predict the series diminishes.

The quest for both the best and worst performing states is also noteworthy, as it aids in determining suitable initial conditions for the reservoir.  We have provided this analysis in \cref{sec:random-states} of the Supplementary Material.

\begin{table*}
    \centering
    \caption{This table presents a summary of simulation results outlined in Section \ref{sec:ParamSpace}. Each cell corresponds to fixed values of $K$ and $\kappa$. All values in the table, including the test error (best, average, and worst) as the first entry, the average quantumness of states with standard deviation as the second entry, and the indices denoting the best and worst performing states in the form (best, worst) as the third entry, are obtained by subjecting 35 random states to the training process under the specified hyperparameters.}
    \label{tab:full-data}
\begin{adjustbox}{max width=\textwidth}
\begin{tabular}{||c|c|c|c|c|c||}
\hline
\multicolumn{1}{||c}{} & $K = 0.02$ & $K = 0.05$ & $K = 0.07$ & $K = 0.1$ & $K=0.12$\\
\hline
$\kappa = 0.02$ & \makecell{$(\textbf{0.0632},0.0658,0.0694)$\\$143.6502 \pm 16.5434$\\$(16,22)$} & \makecell{$(\textbf{0.0592},0.0652,0.0685)$\\$154.5753 \pm 28.1046$\\$(28,16)$} & \makecell{$(\textbf{0.0573},0.0655,0.0810)$\\$162.0891 \pm 16.6570$\\$(20,19)$} & \makecell{$(\textbf{0.0595},0.0705,0.0845)$\\$168.6424 \pm 15.7672$\\$(2,19)$} & \makecell{$(\textbf{0.0639},0.0759,0.0948)$\\$170.5078 \pm 14.8454$\\$(1,13)$}\\
\hline
$\kappa = 0.03$ & \makecell{$(\textbf{0.0633},0.0661,0.0692)$\\$100.9440 \pm 27.6376$\\$(13, 9)$} & \makecell{$(\textbf{0.0605},0.0653,0.0684)$\\$107.8367 \pm 16.6010$\\$(28, 22)$} & \makecell{$(\textbf{0.0579},0.0651,0.0746)$\\$112.7951 \pm 16.6570$\\$(20,19)$} & \makecell{$(\textbf{0.0577},0.0687,0.0832)$\\$117.3331 \pm 15.7672$\\$(2,19)$} & \makecell{$(\textbf{0.0598},0.0728,0.0855)$\\$118.9255 \pm 14.8454$\\$(1,13)$}\\ 
\hline
$\kappa = 0.05$ & \makecell{$(\textbf{0.0634},0.0666,0.0693)$\\$60.8518 \pm 8.6175$\\$(6,22)$} & \makecell{$(\textbf{0.0628},0.0659,0.0685)$\\$64.2209 \pm 16.6010$\\$(28, 22)$} & \makecell{$(\textbf{0.0601},0.0654,0.0713)$\\$67.1316 \pm 16.6570$\\$(20,19)$} & \makecell{$(\textbf{0.0603},0.0674,0.0788)$\\$70.1166 \pm 15.7672$\\$(2,19)$} & \makecell{$(\textbf{0.0590},0.0691,0.790)$\\$71.2287 \pm 14.8454$\\$(1,13)$}\\ 
\hline
$\kappa = 0.1$ & \makecell{$(\textbf{0.0644},0.0671,0.0695)$\\$30.6725 \pm 4.5540$\\$(14,19)$} & \makecell{$(\textbf{0.0628},0.0664,0.0693)$\\$31.5354 \pm 8.6175$\\$(26,12)$} & \makecell{$(\textbf{0.0640},0.0665,0.0693)$\\$32.3377 \pm 8.5992$\\$(7,12)$} & \makecell{$(\textbf{0.0618},0.0663,0.0693)$\\$33.5689 \pm 8.3105$\\$(14,19)$} & \makecell{$(\textbf{0.0628},0.0672,0.0693)$\\$34.1637 \pm 7.9619$\\$(28,19)$}\\
\hline
$\kappa = 0.2$ & \makecell{$(\textbf{0.0658},0.0681,0.0698)$\\$15.9071 \pm 3.2411$\\$(7,22)$} & \makecell{$(\textbf{0.0654},0.0676,0.0693)$\\$16.1002 \pm 4.5540$\\$(14,19)$} & \makecell{$(\textbf{0.0653},0.0674,0.0693)$\\$16.2353 \pm 4.5746$\\$(7,12)$} & \makecell{$(\textbf{0.0647},0.0671,0.0693)$\\$16.4535 \pm 4.5242$\\$(14,19)$} & \makecell{$(\textbf{0.0643},0.0668,0.0698)$\\$16.6154 \pm 4.4510$\\$(28,19)$}\\
\hline
$\kappa = 0.3$ & \makecell{$(\textbf{0.0663},0.0681,0.0698)$\\$11.0579 \pm 3.2411$\\$(27,5)$} & \makecell{$(\textbf{0.0666},0.0676,0.0697)$\\$11.1314 \pm 4.5540$\\$(7,5)$} & \makecell{$(\textbf{0.0660},0.0674,0.0697)$\\$11.1876 \pm 4.5746$\\$(13,5)$} & \makecell{$(\textbf{0.0659},0.0671,0.0699)$\\$11.2706 \pm 4.5242$\\$(26,5)$} & \makecell{$(\textbf{0.0658},0.0668,0.0695)$\\$11.3266 \pm 4.4510$\\$(13,27)$}\\
\hline
\end{tabular}
\end{adjustbox}
\end{table*}

\subsection{Further investigation}
There are other interesting aspects of learning a time series with such a model. Here we discuss some related questions that are addressed in our Supplementary Material. 

One may ask if there are other families of chaotic time series that can be learned by this model. We affirmatively answer this question in \cref{sec:chaos} in the Supplementary Material.
Furthermore, we study the resilience of this model against noise in \cref{SM:noise} of the Supplementary Material, where we observe that the reservoir is surprisingly well-robust against a variety of sources of error.
Finally, we illustrate a few samples of the trajectories of the reservoir, starting from different initial states in \cref{sec:evolution-animation} of the Supplementary Material.

\section{Discussion}\label{sec:discussion}
In this study, we focused on evaluating the effectiveness of a quantum non-linear oscillator in making time-series predictions and examining how quantumness impacts the quantum learning model. We utilize the Lee-Jeong measure \cite{lee2011quantification} as our quantumness metric, which was initially introduced to measure macroscopicity and in this work is shown to be a quantumness measure as well. Through our methodologies, we discovered that quantumness provides us with a broader set of outcomes, and that the best performance observed is achieved at high quantumness, but that high quantumness alone does not guarantee good performance. Overall, our findings contribute to a deeper understanding of the role of quantumness in continuous-variable reservoir computing and highlight its potential for enhancing the performance of this computational model.

Our work raises a number of important questions. Firstly, we aim to determine what specific structures within a reservoir computing model will lead to quantum speed-ups. Indeed, there are other measures of quantumness in the Fock space (c.f., \cite{steuernagel2023quantumness}). It would be interesting to explore the relationship of these measures with learning performance. Additionally, one can investigate the impact of quantumness for a network of oscillators in future research. Notably, when dealing with a network of continuous variable oscillators, entanglement as a measure of quantumness could also be examined. It is worth mentioning that our method has potential for implementation on actual quantum hardware, and may even be feasible with current limited devices due to the strong Kerr non-linearity present in models for a transmon superconducting qubit \cite{bertet2012circuit}. 

\section*{Code Availability}
The codes used for the generation of the plots of this manuscript are publicly available at \href{https://github.com/arsalan-motamedi/QRC}{https://github.com/arsalan-motamedi/QRC}.

\section*{Acknowledgement} We acknowledge Wilten Nicola for fruitful discussions and for reading and commenting on an earlier version of the manuscript. The authors acknowledge an NSERC Discovery Grant, the Alberta Major Innovation Fund, Alberta Innovates, Quantum City, and NRC CSTIP grant AQC 007.  
\section*{Contributions} 
All authors contributed extensively to the presented work. H.Z-H. and C.S. conceived the original ideas and supervised the project. A.M. performed analytical studies and numerical simulations and generated different versions of the manuscript. H.Z-H. and C.S. verified the calculations,  provided detailed feedback on the manuscript, and applied many insightful updates.

\bibliography{main}

\begin{thebibliography}{64}%
\makeatletter
\providecommand \@ifxundefined [1]{%
 \@ifx{#1\undefined}
}%
\providecommand \@ifnum [1]{%
 \ifnum #1\expandafter \@firstoftwo
 \else \expandafter \@secondoftwo
 \fi
}%
\providecommand \@ifx [1]{%
 \ifx #1\expandafter \@firstoftwo
 \else \expandafter \@secondoftwo
 \fi
}%
\providecommand \natexlab [1]{#1}%
\providecommand \enquote  [1]{``#1''}%
\providecommand \bibnamefont  [1]{#1}%
\providecommand \bibfnamefont [1]{#1}%
\providecommand \citenamefont [1]{#1}%
\providecommand \href@noop [0]{\@secondoftwo}%
\providecommand \href [0]{\begingroup \@sanitize@url \@href}%
\providecommand \@href[1]{\@@startlink{#1}\@@href}%
\providecommand \@@href[1]{\endgroup#1\@@endlink}%
\providecommand \@sanitize@url [0]{\catcode `\\12\catcode `\$12\catcode `\&12\catcode `\#12\catcode `\^12\catcode `\_12\catcode `\%12\relax}%
\providecommand \@@startlink[1]{}%
\providecommand \@@endlink[0]{}%
\providecommand \url  [0]{\begingroup\@sanitize@url \@url }%
\providecommand \@url [1]{\endgroup\@href {#1}{\urlprefix }}%
\providecommand \urlprefix  [0]{URL }%
\providecommand \Eprint [0]{\href }%
\providecommand \doibase [0]{https://doi.org/}%
\providecommand \selectlanguage [0]{\@gobble}%
\providecommand \bibinfo  [0]{\@secondoftwo}%
\providecommand \bibfield  [0]{\@secondoftwo}%
\providecommand \translation [1]{[#1]}%
\providecommand \BibitemOpen [0]{}%
\providecommand \bibitemStop [0]{}%
\providecommand \bibitemNoStop [0]{.\EOS\space}%
\providecommand \EOS [0]{\spacefactor3000\relax}%
\providecommand \BibitemShut  [1]{\csname bibitem#1\endcsname}%
\let\auto@bib@innerbib\@empty
\bibitem [{\citenamefont {Shor}(1999)}]{shor1999polynomial}%
  \BibitemOpen
  \bibfield  {author} {\bibinfo {author} {\bibfnamefont {P.~W.}\ \bibnamefont {Shor}},\ }\bibfield  {title} {\bibinfo {title} {Polynomial-time algorithms for prime factorization and discrete logarithms on a quantum computer},\ }\href@noop {} {\bibfield  {journal} {\bibinfo  {journal} {SIAM review}\ }\textbf {\bibinfo {volume} {41}},\ \bibinfo {pages} {303} (\bibinfo {year} {1999})}\BibitemShut {NoStop}%
\bibitem [{\citenamefont {MacQuarrie}\ \emph {et~al.}(2020)\citenamefont {MacQuarrie}, \citenamefont {Simon}, \citenamefont {Simmons},\ and\ \citenamefont {Maine}}]{MacQuarrie_2020}%
  \BibitemOpen
  \bibfield  {author} {\bibinfo {author} {\bibfnamefont {E.~R.}\ \bibnamefont {MacQuarrie}}, \bibinfo {author} {\bibfnamefont {C.}~\bibnamefont {Simon}}, \bibinfo {author} {\bibfnamefont {S.}~\bibnamefont {Simmons}},\ and\ \bibinfo {author} {\bibfnamefont {E.}~\bibnamefont {Maine}},\ }\bibfield  {title} {\bibinfo {title} {The emerging commercial landscape of quantum computing},\ }\href {https://doi.org/10.1038/s42254-020-00247-5} {\bibfield  {journal} {\bibinfo  {journal} {Nature Reviews Physics}\ }\textbf {\bibinfo {volume} {2}},\ \bibinfo {pages} {596} (\bibinfo {year} {2020})}\BibitemShut {NoStop}%
\bibitem [{\citenamefont {Harrow}\ \emph {et~al.}(2009)\citenamefont {Harrow}, \citenamefont {Hassidim},\ and\ \citenamefont {Lloyd}}]{Harrow_2009}%
  \BibitemOpen
  \bibfield  {author} {\bibinfo {author} {\bibfnamefont {A.~W.}\ \bibnamefont {Harrow}}, \bibinfo {author} {\bibfnamefont {A.}~\bibnamefont {Hassidim}},\ and\ \bibinfo {author} {\bibfnamefont {S.}~\bibnamefont {Lloyd}},\ }\bibfield  {title} {\bibinfo {title} {Quantum algorithm for linear systems of equations},\ }\bibfield  {journal} {\bibinfo  {journal} {Physical Review Letters}\ }\textbf {\bibinfo {volume} {103}},\ \href {https://doi.org/10.1103/physrevlett.103.150502} {10.1103/physrevlett.103.150502} (\bibinfo {year} {2009})\BibitemShut {NoStop}%
\bibitem [{\citenamefont {Nielsen}\ and\ \citenamefont {Chuang}(2002)}]{nielsen2002quantum}%
  \BibitemOpen
  \bibfield  {author} {\bibinfo {author} {\bibfnamefont {M.~A.}\ \bibnamefont {Nielsen}}\ and\ \bibinfo {author} {\bibfnamefont {I.}~\bibnamefont {Chuang}},\ }\href@noop {} {\bibinfo {title} {Quantum computation and quantum information}} (\bibinfo {year} {2002})\BibitemShut {NoStop}%
\bibitem [{\citenamefont {Bennett}\ and\ \citenamefont {Brassard}(2020)}]{bennett2020quantum}%
  \BibitemOpen
  \bibfield  {author} {\bibinfo {author} {\bibfnamefont {C.~H.}\ \bibnamefont {Bennett}}\ and\ \bibinfo {author} {\bibfnamefont {G.}~\bibnamefont {Brassard}},\ }\bibfield  {title} {\bibinfo {title} {Quantum cryptography: Public key distribution and coin tossing},\ }\href@noop {} {\bibfield  {journal} {\bibinfo  {journal} {arXiv preprint arXiv:2003.06557}\ } (\bibinfo {year} {2020})}\BibitemShut {NoStop}%
\bibitem [{\citenamefont {Degen}\ \emph {et~al.}(2017)\citenamefont {Degen}, \citenamefont {Reinhard},\ and\ \citenamefont {Cappellaro}}]{degen2017quantum}%
  \BibitemOpen
  \bibfield  {author} {\bibinfo {author} {\bibfnamefont {C.~L.}\ \bibnamefont {Degen}}, \bibinfo {author} {\bibfnamefont {F.}~\bibnamefont {Reinhard}},\ and\ \bibinfo {author} {\bibfnamefont {P.}~\bibnamefont {Cappellaro}},\ }\bibfield  {title} {\bibinfo {title} {Quantum sensing},\ }\href@noop {} {\bibfield  {journal} {\bibinfo  {journal} {Reviews of modern physics}\ }\textbf {\bibinfo {volume} {89}},\ \bibinfo {pages} {035002} (\bibinfo {year} {2017})}\BibitemShut {NoStop}%
\bibitem [{\citenamefont {Simon}(2017)}]{simon2017towards}%
  \BibitemOpen
  \bibfield  {author} {\bibinfo {author} {\bibfnamefont {C.}~\bibnamefont {Simon}},\ }\bibfield  {title} {\bibinfo {title} {Towards a global quantum network},\ }\href@noop {} {\bibfield  {journal} {\bibinfo  {journal} {Nature Photonics}\ }\textbf {\bibinfo {volume} {11}},\ \bibinfo {pages} {678} (\bibinfo {year} {2017})}\BibitemShut {NoStop}%
\bibitem [{\citenamefont {Rivest}\ \emph {et~al.}(1983)\citenamefont {Rivest}, \citenamefont {Shamir},\ and\ \citenamefont {Adleman}}]{rivest1983cryptographic}%
  \BibitemOpen
  \bibfield  {author} {\bibinfo {author} {\bibfnamefont {R.~L.}\ \bibnamefont {Rivest}}, \bibinfo {author} {\bibfnamefont {A.}~\bibnamefont {Shamir}},\ and\ \bibinfo {author} {\bibfnamefont {L.~M.}\ \bibnamefont {Adleman}},\ }\href@noop {} {\bibinfo {title} {Cryptographic communications system and method}} (\bibinfo {year} {1983}),\ \bibinfo {note} {uS Patent 4,405,829}\BibitemShut {NoStop}%
\bibitem [{\citenamefont {Aharonov}\ and\ \citenamefont {Ben-Or}(1997)}]{aharonov1997fault}%
  \BibitemOpen
  \bibfield  {author} {\bibinfo {author} {\bibfnamefont {D.}~\bibnamefont {Aharonov}}\ and\ \bibinfo {author} {\bibfnamefont {M.}~\bibnamefont {Ben-Or}},\ }\bibfield  {title} {\bibinfo {title} {Fault-tolerant quantum computation with constant error},\ }in\ \href@noop {} {\emph {\bibinfo {booktitle} {Proceedings of the twenty-ninth annual ACM symposium on Theory of computing}}}\ (\bibinfo {year} {1997})\ pp.\ \bibinfo {pages} {176--188}\BibitemShut {NoStop}%
\bibitem [{\citenamefont {Knill}\ \emph {et~al.}(1998)\citenamefont {Knill}, \citenamefont {Laflamme},\ and\ \citenamefont {Zurek}}]{knill1998resilient}%
  \BibitemOpen
  \bibfield  {author} {\bibinfo {author} {\bibfnamefont {E.}~\bibnamefont {Knill}}, \bibinfo {author} {\bibfnamefont {R.}~\bibnamefont {Laflamme}},\ and\ \bibinfo {author} {\bibfnamefont {W.~H.}\ \bibnamefont {Zurek}},\ }\bibfield  {title} {\bibinfo {title} {Resilient quantum computation},\ }\href@noop {} {\bibfield  {journal} {\bibinfo  {journal} {Science}\ }\textbf {\bibinfo {volume} {279}},\ \bibinfo {pages} {342} (\bibinfo {year} {1998})}\BibitemShut {NoStop}%
\bibitem [{\citenamefont {Kitaev}(2003)}]{kitaev2003fault}%
  \BibitemOpen
  \bibfield  {author} {\bibinfo {author} {\bibfnamefont {A.~Y.}\ \bibnamefont {Kitaev}},\ }\bibfield  {title} {\bibinfo {title} {Fault-tolerant quantum computation by anyons},\ }\href@noop {} {\bibfield  {journal} {\bibinfo  {journal} {Annals of Physics}\ }\textbf {\bibinfo {volume} {303}},\ \bibinfo {pages} {2} (\bibinfo {year} {2003})}\BibitemShut {NoStop}%
\bibitem [{\citenamefont {Shor}(1996)}]{shor1996fault}%
  \BibitemOpen
  \bibfield  {author} {\bibinfo {author} {\bibfnamefont {P.~W.}\ \bibnamefont {Shor}},\ }\bibfield  {title} {\bibinfo {title} {Fault-tolerant quantum computation},\ }in\ \href@noop {} {\emph {\bibinfo {booktitle} {Proceedings of 37th conference on foundations of computer science}}}\ (\bibinfo {organization} {IEEE},\ \bibinfo {year} {1996})\ pp.\ \bibinfo {pages} {56--65}\BibitemShut {NoStop}%
\bibitem [{\citenamefont {Acharya}\ \emph {et~al.}(2022)\citenamefont {Acharya}, \citenamefont {Aleiner}, \citenamefont {Allen}, \citenamefont {Andersen}, \citenamefont {Ansmann}, \citenamefont {Arute}, \citenamefont {Arya}, \citenamefont {Asfaw}, \citenamefont {Atalaya}, \citenamefont {Babbush} \emph {et~al.}}]{acharya2022suppressing}%
  \BibitemOpen
  \bibfield  {author} {\bibinfo {author} {\bibfnamefont {R.}~\bibnamefont {Acharya}}, \bibinfo {author} {\bibfnamefont {I.}~\bibnamefont {Aleiner}}, \bibinfo {author} {\bibfnamefont {R.}~\bibnamefont {Allen}}, \bibinfo {author} {\bibfnamefont {T.~I.}\ \bibnamefont {Andersen}}, \bibinfo {author} {\bibfnamefont {M.}~\bibnamefont {Ansmann}}, \bibinfo {author} {\bibfnamefont {F.}~\bibnamefont {Arute}}, \bibinfo {author} {\bibfnamefont {K.}~\bibnamefont {Arya}}, \bibinfo {author} {\bibfnamefont {A.}~\bibnamefont {Asfaw}}, \bibinfo {author} {\bibfnamefont {J.}~\bibnamefont {Atalaya}}, \bibinfo {author} {\bibfnamefont {R.}~\bibnamefont {Babbush}}, \emph {et~al.},\ }\bibfield  {title} {\bibinfo {title} {Suppressing quantum errors by scaling a surface code logical qubit},\ }\href@noop {} {\bibfield  {journal} {\bibinfo  {journal} {arXiv preprint arXiv:2207.06431}\ } (\bibinfo {year} {2022})}\BibitemShut {NoStop}%
\bibitem [{\citenamefont {Temme}\ \emph {et~al.}(2017)\citenamefont {Temme}, \citenamefont {Bravyi},\ and\ \citenamefont {Gambetta}}]{temme2017error}%
  \BibitemOpen
  \bibfield  {author} {\bibinfo {author} {\bibfnamefont {K.}~\bibnamefont {Temme}}, \bibinfo {author} {\bibfnamefont {S.}~\bibnamefont {Bravyi}},\ and\ \bibinfo {author} {\bibfnamefont {J.~M.}\ \bibnamefont {Gambetta}},\ }\bibfield  {title} {\bibinfo {title} {Error mitigation for short-depth quantum circuits},\ }\href@noop {} {\bibfield  {journal} {\bibinfo  {journal} {Physical review letters}\ }\textbf {\bibinfo {volume} {119}},\ \bibinfo {pages} {180509} (\bibinfo {year} {2017})}\BibitemShut {NoStop}%
\bibitem [{\citenamefont {Bharti}\ \emph {et~al.}(2022)\citenamefont {Bharti}, \citenamefont {Cervera-Lierta}, \citenamefont {Kyaw}, \citenamefont {Haug}, \citenamefont {Alperin-Lea}, \citenamefont {Anand}, \citenamefont {Degroote}, \citenamefont {Heimonen}, \citenamefont {Kottmann}, \citenamefont {Menke} \emph {et~al.}}]{bharti2022noisy}%
  \BibitemOpen
  \bibfield  {author} {\bibinfo {author} {\bibfnamefont {K.}~\bibnamefont {Bharti}}, \bibinfo {author} {\bibfnamefont {A.}~\bibnamefont {Cervera-Lierta}}, \bibinfo {author} {\bibfnamefont {T.~H.}\ \bibnamefont {Kyaw}}, \bibinfo {author} {\bibfnamefont {T.}~\bibnamefont {Haug}}, \bibinfo {author} {\bibfnamefont {S.}~\bibnamefont {Alperin-Lea}}, \bibinfo {author} {\bibfnamefont {A.}~\bibnamefont {Anand}}, \bibinfo {author} {\bibfnamefont {M.}~\bibnamefont {Degroote}}, \bibinfo {author} {\bibfnamefont {H.}~\bibnamefont {Heimonen}}, \bibinfo {author} {\bibfnamefont {J.~S.}\ \bibnamefont {Kottmann}}, \bibinfo {author} {\bibfnamefont {T.}~\bibnamefont {Menke}}, \emph {et~al.},\ }\bibfield  {title} {\bibinfo {title} {Noisy intermediate-scale quantum algorithms},\ }\href@noop {} {\bibfield  {journal} {\bibinfo  {journal} {Reviews of Modern Physics}\ }\textbf {\bibinfo {volume} {94}},\ \bibinfo {pages} {015004} (\bibinfo {year} {2022})}\BibitemShut {NoStop}%
\bibitem [{\citenamefont {Kandala}\ \emph {et~al.}(2019)\citenamefont {Kandala}, \citenamefont {Temme}, \citenamefont {C{\'o}rcoles}, \citenamefont {Mezzacapo}, \citenamefont {Chow},\ and\ \citenamefont {Gambetta}}]{kandala2019error}%
  \BibitemOpen
  \bibfield  {author} {\bibinfo {author} {\bibfnamefont {A.}~\bibnamefont {Kandala}}, \bibinfo {author} {\bibfnamefont {K.}~\bibnamefont {Temme}}, \bibinfo {author} {\bibfnamefont {A.~D.}\ \bibnamefont {C{\'o}rcoles}}, \bibinfo {author} {\bibfnamefont {A.}~\bibnamefont {Mezzacapo}}, \bibinfo {author} {\bibfnamefont {J.~M.}\ \bibnamefont {Chow}},\ and\ \bibinfo {author} {\bibfnamefont {J.~M.}\ \bibnamefont {Gambetta}},\ }\bibfield  {title} {\bibinfo {title} {Error mitigation extends the computational reach of a noisy quantum processor},\ }\href@noop {} {\bibfield  {journal} {\bibinfo  {journal} {Nature}\ }\textbf {\bibinfo {volume} {567}},\ \bibinfo {pages} {491} (\bibinfo {year} {2019})}\BibitemShut {NoStop}%
\bibitem [{\citenamefont {Preskill}(2018)}]{preskill2018quantum}%
  \BibitemOpen
  \bibfield  {author} {\bibinfo {author} {\bibfnamefont {J.}~\bibnamefont {Preskill}},\ }\bibfield  {title} {\bibinfo {title} {Quantum computing in the nisq era and beyond},\ }\href@noop {} {\bibfield  {journal} {\bibinfo  {journal} {Quantum}\ }\textbf {\bibinfo {volume} {2}},\ \bibinfo {pages} {79} (\bibinfo {year} {2018})}\BibitemShut {NoStop}%
\bibitem [{\citenamefont {Farquhar}\ \emph {et~al.}(2006)\citenamefont {Farquhar}, \citenamefont {Gordon},\ and\ \citenamefont {Hasler}}]{farquhar2006field}%
  \BibitemOpen
  \bibfield  {author} {\bibinfo {author} {\bibfnamefont {E.}~\bibnamefont {Farquhar}}, \bibinfo {author} {\bibfnamefont {C.}~\bibnamefont {Gordon}},\ and\ \bibinfo {author} {\bibfnamefont {P.}~\bibnamefont {Hasler}},\ }\bibfield  {title} {\bibinfo {title} {A field programmable neural array},\ }in\ \href@noop {} {\emph {\bibinfo {booktitle} {2006 IEEE international symposium on circuits and systems}}}\ (\bibinfo {organization} {IEEE},\ \bibinfo {year} {2006})\ pp.\ \bibinfo {pages} {4--pp}\BibitemShut {NoStop}%
\bibitem [{\citenamefont {Hopfield}(1982)}]{hopfield1982neural}%
  \BibitemOpen
  \bibfield  {author} {\bibinfo {author} {\bibfnamefont {J.~J.}\ \bibnamefont {Hopfield}},\ }\bibfield  {title} {\bibinfo {title} {Neural networks and physical systems with emergent collective computational abilities.},\ }\href@noop {} {\bibfield  {journal} {\bibinfo  {journal} {Proceedings of the national academy of sciences}\ }\textbf {\bibinfo {volume} {79}},\ \bibinfo {pages} {2554} (\bibinfo {year} {1982})}\BibitemShut {NoStop}%
\bibitem [{\citenamefont {Schmidhuber}(2015)}]{schmidhuber2015deep}%
  \BibitemOpen
  \bibfield  {author} {\bibinfo {author} {\bibfnamefont {J.}~\bibnamefont {Schmidhuber}},\ }\bibfield  {title} {\bibinfo {title} {Deep learning in neural networks: An overview},\ }\href@noop {} {\bibfield  {journal} {\bibinfo  {journal} {Neural networks}\ }\textbf {\bibinfo {volume} {61}},\ \bibinfo {pages} {85} (\bibinfo {year} {2015})}\BibitemShut {NoStop}%
\bibitem [{\citenamefont {Goodfellow}\ \emph {et~al.}(2020)\citenamefont {Goodfellow}, \citenamefont {Pouget-Abadie}, \citenamefont {Mirza}, \citenamefont {Xu}, \citenamefont {Warde-Farley}, \citenamefont {Ozair}, \citenamefont {Courville},\ and\ \citenamefont {Bengio}}]{goodfellow2020generative}%
  \BibitemOpen
  \bibfield  {author} {\bibinfo {author} {\bibfnamefont {I.}~\bibnamefont {Goodfellow}}, \bibinfo {author} {\bibfnamefont {J.}~\bibnamefont {Pouget-Abadie}}, \bibinfo {author} {\bibfnamefont {M.}~\bibnamefont {Mirza}}, \bibinfo {author} {\bibfnamefont {B.}~\bibnamefont {Xu}}, \bibinfo {author} {\bibfnamefont {D.}~\bibnamefont {Warde-Farley}}, \bibinfo {author} {\bibfnamefont {S.}~\bibnamefont {Ozair}}, \bibinfo {author} {\bibfnamefont {A.}~\bibnamefont {Courville}},\ and\ \bibinfo {author} {\bibfnamefont {Y.}~\bibnamefont {Bengio}},\ }\bibfield  {title} {\bibinfo {title} {Generative adversarial networks},\ }\href@noop {} {\bibfield  {journal} {\bibinfo  {journal} {Communications of the ACM}\ }\textbf {\bibinfo {volume} {63}},\ \bibinfo {pages} {139} (\bibinfo {year} {2020})}\BibitemShut {NoStop}%
\bibitem [{\citenamefont {Pascanu}\ \emph {et~al.}(2013)\citenamefont {Pascanu}, \citenamefont {Mikolov},\ and\ \citenamefont {Bengio}}]{pascanu2013difficulty}%
  \BibitemOpen
  \bibfield  {author} {\bibinfo {author} {\bibfnamefont {R.}~\bibnamefont {Pascanu}}, \bibinfo {author} {\bibfnamefont {T.}~\bibnamefont {Mikolov}},\ and\ \bibinfo {author} {\bibfnamefont {Y.}~\bibnamefont {Bengio}},\ }\bibfield  {title} {\bibinfo {title} {On the difficulty of training recurrent neural networks},\ }in\ \href@noop {} {\emph {\bibinfo {booktitle} {International conference on machine learning}}}\ (\bibinfo {organization} {PMLR},\ \bibinfo {year} {2013})\ pp.\ \bibinfo {pages} {1310--1318}\BibitemShut {NoStop}%
\bibitem [{\citenamefont {Basodi}\ \emph {et~al.}(2020)\citenamefont {Basodi}, \citenamefont {Ji}, \citenamefont {Zhang},\ and\ \citenamefont {Pan}}]{basodi2020gradient}%
  \BibitemOpen
  \bibfield  {author} {\bibinfo {author} {\bibfnamefont {S.}~\bibnamefont {Basodi}}, \bibinfo {author} {\bibfnamefont {C.}~\bibnamefont {Ji}}, \bibinfo {author} {\bibfnamefont {H.}~\bibnamefont {Zhang}},\ and\ \bibinfo {author} {\bibfnamefont {Y.}~\bibnamefont {Pan}},\ }\bibfield  {title} {\bibinfo {title} {Gradient amplification: An efficient way to train deep neural networks},\ }\href@noop {} {\bibfield  {journal} {\bibinfo  {journal} {Big Data Mining and Analytics}\ }\textbf {\bibinfo {volume} {3}},\ \bibinfo {pages} {196} (\bibinfo {year} {2020})}\BibitemShut {NoStop}%
\bibitem [{\citenamefont {Maass}\ \emph {et~al.}(2002)\citenamefont {Maass}, \citenamefont {Natschl{\"a}ger},\ and\ \citenamefont {Markram}}]{maass2002real}%
  \BibitemOpen
  \bibfield  {author} {\bibinfo {author} {\bibfnamefont {W.}~\bibnamefont {Maass}}, \bibinfo {author} {\bibfnamefont {T.}~\bibnamefont {Natschl{\"a}ger}},\ and\ \bibinfo {author} {\bibfnamefont {H.}~\bibnamefont {Markram}},\ }\bibfield  {title} {\bibinfo {title} {Real-time computing without stable states: A new framework for neural computation based on perturbations},\ }\href@noop {} {\bibfield  {journal} {\bibinfo  {journal} {Neural computation}\ }\textbf {\bibinfo {volume} {14}},\ \bibinfo {pages} {2531} (\bibinfo {year} {2002})}\BibitemShut {NoStop}%
\bibitem [{\citenamefont {Jaeger}\ and\ \citenamefont {Haas}(2004)}]{jaeger2004harnessing}%
  \BibitemOpen
  \bibfield  {author} {\bibinfo {author} {\bibfnamefont {H.}~\bibnamefont {Jaeger}}\ and\ \bibinfo {author} {\bibfnamefont {H.}~\bibnamefont {Haas}},\ }\bibfield  {title} {\bibinfo {title} {Harnessing nonlinearity: Predicting chaotic systems and saving energy in wireless communication},\ }\href@noop {} {\bibfield  {journal} {\bibinfo  {journal} {science}\ }\textbf {\bibinfo {volume} {304}},\ \bibinfo {pages} {78} (\bibinfo {year} {2004})}\BibitemShut {NoStop}%
\bibitem [{\citenamefont {Tanaka}\ \emph {et~al.}(2019)\citenamefont {Tanaka}, \citenamefont {Yamane}, \citenamefont {H{\'e}roux}, \citenamefont {Nakane}, \citenamefont {Kanazawa}, \citenamefont {Takeda}, \citenamefont {Numata}, \citenamefont {Nakano},\ and\ \citenamefont {Hirose}}]{tanaka2019recent}%
  \BibitemOpen
  \bibfield  {author} {\bibinfo {author} {\bibfnamefont {G.}~\bibnamefont {Tanaka}}, \bibinfo {author} {\bibfnamefont {T.}~\bibnamefont {Yamane}}, \bibinfo {author} {\bibfnamefont {J.~B.}\ \bibnamefont {H{\'e}roux}}, \bibinfo {author} {\bibfnamefont {R.}~\bibnamefont {Nakane}}, \bibinfo {author} {\bibfnamefont {N.}~\bibnamefont {Kanazawa}}, \bibinfo {author} {\bibfnamefont {S.}~\bibnamefont {Takeda}}, \bibinfo {author} {\bibfnamefont {H.}~\bibnamefont {Numata}}, \bibinfo {author} {\bibfnamefont {D.}~\bibnamefont {Nakano}},\ and\ \bibinfo {author} {\bibfnamefont {A.}~\bibnamefont {Hirose}},\ }\bibfield  {title} {\bibinfo {title} {Recent advances in physical reservoir computing: A review},\ }\href@noop {} {\bibfield  {journal} {\bibinfo  {journal} {Neural Networks}\ }\textbf {\bibinfo {volume} {115}},\ \bibinfo {pages} {100} (\bibinfo {year} {2019})}\BibitemShut {NoStop}%
\bibitem [{\citenamefont {R{\"o}hm}\ and\ \citenamefont {L{\"u}dge}(2018)}]{rohm2018multiplexed}%
  \BibitemOpen
  \bibfield  {author} {\bibinfo {author} {\bibfnamefont {A.}~\bibnamefont {R{\"o}hm}}\ and\ \bibinfo {author} {\bibfnamefont {K.}~\bibnamefont {L{\"u}dge}},\ }\bibfield  {title} {\bibinfo {title} {Multiplexed networks: reservoir computing with virtual and real nodes},\ }\href@noop {} {\bibfield  {journal} {\bibinfo  {journal} {Journal of Physics Communications}\ }\textbf {\bibinfo {volume} {2}},\ \bibinfo {pages} {085007} (\bibinfo {year} {2018})}\BibitemShut {NoStop}%
\bibitem [{\citenamefont {Nicola}\ and\ \citenamefont {Clopath}(2017)}]{nature1}%
  \BibitemOpen
  \bibfield  {author} {\bibinfo {author} {\bibfnamefont {W.}~\bibnamefont {Nicola}}\ and\ \bibinfo {author} {\bibfnamefont {C.}~\bibnamefont {Clopath}},\ }\bibfield  {title} {\bibinfo {title} {Supervised learning in spiking neural networks with force training},\ }\bibfield  {journal} {\bibinfo  {journal} {Nat Commun}\ }\textbf {\bibinfo {volume} {8}},\ \href {https://doi.org/10.1038/s41467-017-01827-3} {10.1038/s41467-017-01827-3} (\bibinfo {year} {2017})\BibitemShut {NoStop}%
\bibitem [{\citenamefont {Nakajima}(2018)}]{nakajima2018reservoir}%
  \BibitemOpen
  \bibfield  {author} {\bibinfo {author} {\bibfnamefont {K.}~\bibnamefont {Nakajima}},\ }\bibfield  {title} {\bibinfo {title} {Reservoir computing: Theory, physical implementations, and applications},\ }\href@noop {} {\bibfield  {journal} {\bibinfo  {journal} {IEICE Technical Report; IEICE Tech. Rep.}\ }\textbf {\bibinfo {volume} {118}},\ \bibinfo {pages} {149} (\bibinfo {year} {2018})}\BibitemShut {NoStop}%
\bibitem [{\citenamefont {Schrauwen}\ \emph {et~al.}(2007)\citenamefont {Schrauwen}, \citenamefont {Verstraeten},\ and\ \citenamefont {Van~Campenhout}}]{schrauwen2007overview}%
  \BibitemOpen
  \bibfield  {author} {\bibinfo {author} {\bibfnamefont {B.}~\bibnamefont {Schrauwen}}, \bibinfo {author} {\bibfnamefont {D.}~\bibnamefont {Verstraeten}},\ and\ \bibinfo {author} {\bibfnamefont {J.}~\bibnamefont {Van~Campenhout}},\ }\bibfield  {title} {\bibinfo {title} {An overview of reservoir computing: theory, applications and implementations},\ }in\ \href@noop {} {\emph {\bibinfo {booktitle} {Proceedings of the 15th european symposium on artificial neural networks. p. 471-482 2007}}}\ (\bibinfo {year} {2007})\ pp.\ \bibinfo {pages} {471--482}\BibitemShut {NoStop}%
\bibitem [{\citenamefont {Mammedov}\ \emph {et~al.}(2022)\citenamefont {Mammedov}, \citenamefont {Olugu},\ and\ \citenamefont {Farah}}]{mammedov2022weather}%
  \BibitemOpen
  \bibfield  {author} {\bibinfo {author} {\bibfnamefont {Y.~D.}\ \bibnamefont {Mammedov}}, \bibinfo {author} {\bibfnamefont {E.~U.}\ \bibnamefont {Olugu}},\ and\ \bibinfo {author} {\bibfnamefont {G.~A.}\ \bibnamefont {Farah}},\ }\bibfield  {title} {\bibinfo {title} {Weather forecasting based on data-driven and physics-informed reservoir computing models},\ }\href@noop {} {\bibfield  {journal} {\bibinfo  {journal} {Environmental Science and Pollution Research}\ }\textbf {\bibinfo {volume} {29}},\ \bibinfo {pages} {24131} (\bibinfo {year} {2022})}\BibitemShut {NoStop}%
\bibitem [{\citenamefont {Kan}\ \emph {et~al.}(2022)\citenamefont {Kan}, \citenamefont {Nakajima}, \citenamefont {Asai},\ and\ \citenamefont {Akai-Kasaya}}]{kan2022physical}%
  \BibitemOpen
  \bibfield  {author} {\bibinfo {author} {\bibfnamefont {S.}~\bibnamefont {Kan}}, \bibinfo {author} {\bibfnamefont {K.}~\bibnamefont {Nakajima}}, \bibinfo {author} {\bibfnamefont {T.}~\bibnamefont {Asai}},\ and\ \bibinfo {author} {\bibfnamefont {M.}~\bibnamefont {Akai-Kasaya}},\ }\bibfield  {title} {\bibinfo {title} {Physical implementation of reservoir computing through electrochemical reaction},\ }\href@noop {} {\bibfield  {journal} {\bibinfo  {journal} {Advanced Science}\ }\textbf {\bibinfo {volume} {9}},\ \bibinfo {pages} {2104076} (\bibinfo {year} {2022})}\BibitemShut {NoStop}%
\bibitem [{\citenamefont {Biamonte}\ \emph {et~al.}(2017)\citenamefont {Biamonte}, \citenamefont {Wittek}, \citenamefont {Pancotti}, \citenamefont {Rebentrost}, \citenamefont {Wiebe},\ and\ \citenamefont {Lloyd}}]{biamonte2017quantum}%
  \BibitemOpen
  \bibfield  {author} {\bibinfo {author} {\bibfnamefont {J.}~\bibnamefont {Biamonte}}, \bibinfo {author} {\bibfnamefont {P.}~\bibnamefont {Wittek}}, \bibinfo {author} {\bibfnamefont {N.}~\bibnamefont {Pancotti}}, \bibinfo {author} {\bibfnamefont {P.}~\bibnamefont {Rebentrost}}, \bibinfo {author} {\bibfnamefont {N.}~\bibnamefont {Wiebe}},\ and\ \bibinfo {author} {\bibfnamefont {S.}~\bibnamefont {Lloyd}},\ }\bibfield  {title} {\bibinfo {title} {Quantum machine learning},\ }\href@noop {} {\bibfield  {journal} {\bibinfo  {journal} {Nature}\ }\textbf {\bibinfo {volume} {549}},\ \bibinfo {pages} {195} (\bibinfo {year} {2017})}\BibitemShut {NoStop}%
\bibitem [{\citenamefont {Maria~Schuld}()}]{QML}%
  \BibitemOpen
  \bibfield  {author} {\bibinfo {author} {\bibfnamefont {F.~P.}\ \bibnamefont {Maria~Schuld}, \bibfnamefont {Ilya~Sinayskiy}},\ }\bibfield  {title} {\bibinfo {title} {An introduction to quantum machine learning},\ }\bibfield  {journal} {\bibinfo  {journal} {Contemporary Physics}\ }\href {https://doi.org/10.1080/00107514.2014.964942} {10.1080/00107514.2014.964942}\BibitemShut {NoStop}%
\bibitem [{\citenamefont {Schuld}\ \emph {et~al.}(2015)\citenamefont {Schuld}, \citenamefont {Sinayskiy},\ and\ \citenamefont {Petruccione}}]{schuld2015introduction}%
  \BibitemOpen
  \bibfield  {author} {\bibinfo {author} {\bibfnamefont {M.}~\bibnamefont {Schuld}}, \bibinfo {author} {\bibfnamefont {I.}~\bibnamefont {Sinayskiy}},\ and\ \bibinfo {author} {\bibfnamefont {F.}~\bibnamefont {Petruccione}},\ }\bibfield  {title} {\bibinfo {title} {An introduction to quantum machine learning},\ }\href@noop {} {\bibfield  {journal} {\bibinfo  {journal} {Contemporary Physics}\ }\textbf {\bibinfo {volume} {56}},\ \bibinfo {pages} {172} (\bibinfo {year} {2015})}\BibitemShut {NoStop}%
\bibitem [{\citenamefont {Fujii}\ and\ \citenamefont {Nakajima}(2021)}]{fujii2021quantum}%
  \BibitemOpen
  \bibfield  {author} {\bibinfo {author} {\bibfnamefont {K.}~\bibnamefont {Fujii}}\ and\ \bibinfo {author} {\bibfnamefont {K.}~\bibnamefont {Nakajima}},\ }\bibfield  {title} {\bibinfo {title} {Quantum reservoir computing: a reservoir approach toward quantum machine learning on near-term quantum devices},\ }in\ \href@noop {} {\emph {\bibinfo {booktitle} {Reservoir computing}}}\ (\bibinfo  {publisher} {Springer},\ \bibinfo {year} {2021})\ pp.\ \bibinfo {pages} {423--450}\BibitemShut {NoStop}%
\bibitem [{\citenamefont {Govia}\ \emph {et~al.}(2021)\citenamefont {Govia}, \citenamefont {Ribeill}, \citenamefont {Rowlands}, \citenamefont {Krovi},\ and\ \citenamefont {Ohki}}]{PhysRevResearch.3.013077}%
  \BibitemOpen
  \bibfield  {author} {\bibinfo {author} {\bibfnamefont {L.~C.~G.}\ \bibnamefont {Govia}}, \bibinfo {author} {\bibfnamefont {G.~J.}\ \bibnamefont {Ribeill}}, \bibinfo {author} {\bibfnamefont {G.~E.}\ \bibnamefont {Rowlands}}, \bibinfo {author} {\bibfnamefont {H.~K.}\ \bibnamefont {Krovi}},\ and\ \bibinfo {author} {\bibfnamefont {T.~A.}\ \bibnamefont {Ohki}},\ }\bibfield  {title} {\bibinfo {title} {Quantum reservoir computing with a single nonlinear oscillator},\ }\href {https://doi.org/10.1103/PhysRevResearch.3.013077} {\bibfield  {journal} {\bibinfo  {journal} {Phys. Rev. Research}\ }\textbf {\bibinfo {volume} {3}},\ \bibinfo {pages} {013077} (\bibinfo {year} {2021})}\BibitemShut {NoStop}%
\bibitem [{\citenamefont {Luchnikov}\ \emph {et~al.}(2019)\citenamefont {Luchnikov}, \citenamefont {Vintskevich}, \citenamefont {Ouerdane},\ and\ \citenamefont {Filippov}}]{luchnikov2019simulation}%
  \BibitemOpen
  \bibfield  {author} {\bibinfo {author} {\bibfnamefont {I.}~\bibnamefont {Luchnikov}}, \bibinfo {author} {\bibfnamefont {S.}~\bibnamefont {Vintskevich}}, \bibinfo {author} {\bibfnamefont {H.}~\bibnamefont {Ouerdane}},\ and\ \bibinfo {author} {\bibfnamefont {S.}~\bibnamefont {Filippov}},\ }\bibfield  {title} {\bibinfo {title} {Simulation complexity of open quantum dynamics: Connection with tensor networks},\ }\href@noop {} {\bibfield  {journal} {\bibinfo  {journal} {Physical review letters}\ }\textbf {\bibinfo {volume} {122}},\ \bibinfo {pages} {160401} (\bibinfo {year} {2019})}\BibitemShut {NoStop}%
\bibitem [{\citenamefont {Mart{\'\i}nez-Pe{\~n}a}\ \emph {et~al.}(2020)\citenamefont {Mart{\'\i}nez-Pe{\~n}a}, \citenamefont {Nokkala}, \citenamefont {Giorgi}, \citenamefont {Zambrini},\ and\ \citenamefont {Soriano}}]{martinez2020information}%
  \BibitemOpen
  \bibfield  {author} {\bibinfo {author} {\bibfnamefont {R.}~\bibnamefont {Mart{\'\i}nez-Pe{\~n}a}}, \bibinfo {author} {\bibfnamefont {J.}~\bibnamefont {Nokkala}}, \bibinfo {author} {\bibfnamefont {G.~L.}\ \bibnamefont {Giorgi}}, \bibinfo {author} {\bibfnamefont {R.}~\bibnamefont {Zambrini}},\ and\ \bibinfo {author} {\bibfnamefont {M.~C.}\ \bibnamefont {Soriano}},\ }\bibfield  {title} {\bibinfo {title} {Information processing capacity of spin-based quantum reservoir computing systems},\ }\href@noop {} {\bibfield  {journal} {\bibinfo  {journal} {Cognitive Computation}\ ,\ \bibinfo {pages} {1}} (\bibinfo {year} {2020})}\BibitemShut {NoStop}%
\bibitem [{\citenamefont {Nokkala}\ \emph {et~al.}(2021)\citenamefont {Nokkala}, \citenamefont {Mart{\'\i}nez-Pe{\~n}a}, \citenamefont {Giorgi}, \citenamefont {Parigi}, \citenamefont {Soriano},\ and\ \citenamefont {Zambrini}}]{nokkala2021gaussian}%
  \BibitemOpen
  \bibfield  {author} {\bibinfo {author} {\bibfnamefont {J.}~\bibnamefont {Nokkala}}, \bibinfo {author} {\bibfnamefont {R.}~\bibnamefont {Mart{\'\i}nez-Pe{\~n}a}}, \bibinfo {author} {\bibfnamefont {G.~L.}\ \bibnamefont {Giorgi}}, \bibinfo {author} {\bibfnamefont {V.}~\bibnamefont {Parigi}}, \bibinfo {author} {\bibfnamefont {M.~C.}\ \bibnamefont {Soriano}},\ and\ \bibinfo {author} {\bibfnamefont {R.}~\bibnamefont {Zambrini}},\ }\bibfield  {title} {\bibinfo {title} {Gaussian states of continuous-variable quantum systems provide universal and versatile reservoir computing},\ }\href@noop {} {\bibfield  {journal} {\bibinfo  {journal} {Communications Physics}\ }\textbf {\bibinfo {volume} {4}},\ \bibinfo {pages} {53} (\bibinfo {year} {2021})}\BibitemShut {NoStop}%
\bibitem [{\citenamefont {Pfeffer}\ \emph {et~al.}(2022)\citenamefont {Pfeffer}, \citenamefont {Heyder},\ and\ \citenamefont {Schumacher}}]{pfeffer2022quantum}%
  \BibitemOpen
  \bibfield  {author} {\bibinfo {author} {\bibfnamefont {P.}~\bibnamefont {Pfeffer}}, \bibinfo {author} {\bibfnamefont {F.}~\bibnamefont {Heyder}},\ and\ \bibinfo {author} {\bibfnamefont {J.}~\bibnamefont {Schumacher}},\ }\bibfield  {title} {\bibinfo {title} {Quantum reservoir computing of thermal convection flow},\ }\href@noop {} {\bibfield  {journal} {\bibinfo  {journal} {arXiv preprint arXiv:2204.13951}\ } (\bibinfo {year} {2022})}\BibitemShut {NoStop}%
\bibitem [{\citenamefont {Ghosh}\ \emph {et~al.}(2021)\citenamefont {Ghosh}, \citenamefont {Nakajima}, \citenamefont {Krisnanda}, \citenamefont {Fujii},\ and\ \citenamefont {Liew}}]{ghosh2021quantum}%
  \BibitemOpen
  \bibfield  {author} {\bibinfo {author} {\bibfnamefont {S.}~\bibnamefont {Ghosh}}, \bibinfo {author} {\bibfnamefont {K.}~\bibnamefont {Nakajima}}, \bibinfo {author} {\bibfnamefont {T.}~\bibnamefont {Krisnanda}}, \bibinfo {author} {\bibfnamefont {K.}~\bibnamefont {Fujii}},\ and\ \bibinfo {author} {\bibfnamefont {T.~C.}\ \bibnamefont {Liew}},\ }\bibfield  {title} {\bibinfo {title} {Quantum neuromorphic computing with reservoir computing networks},\ }\href@noop {} {\bibfield  {journal} {\bibinfo  {journal} {Advanced Quantum Technologies}\ }\textbf {\bibinfo {volume} {4}},\ \bibinfo {pages} {2100053} (\bibinfo {year} {2021})}\BibitemShut {NoStop}%
\bibitem [{\citenamefont {Vintskevich}\ and\ \citenamefont {Grigoriev}(2022)}]{vintskevich2022computing}%
  \BibitemOpen
  \bibfield  {author} {\bibinfo {author} {\bibfnamefont {S.}~\bibnamefont {Vintskevich}}\ and\ \bibinfo {author} {\bibfnamefont {D.}~\bibnamefont {Grigoriev}},\ }\bibfield  {title} {\bibinfo {title} {Computing with two quantum reservoirs connected via optimized two-qubit nonselective measurements},\ }\href@noop {} {\bibfield  {journal} {\bibinfo  {journal} {arXiv preprint arXiv:2201.07969}\ } (\bibinfo {year} {2022})}\BibitemShut {NoStop}%
\bibitem [{\citenamefont {G{\"o}tting}\ \emph {et~al.}(2023)\citenamefont {G{\"o}tting}, \citenamefont {Lohof},\ and\ \citenamefont {Gies}}]{gotting2023exploring}%
  \BibitemOpen
  \bibfield  {author} {\bibinfo {author} {\bibfnamefont {N.}~\bibnamefont {G{\"o}tting}}, \bibinfo {author} {\bibfnamefont {F.}~\bibnamefont {Lohof}},\ and\ \bibinfo {author} {\bibfnamefont {C.}~\bibnamefont {Gies}},\ }\bibfield  {title} {\bibinfo {title} {Exploring quantum mechanical advantage for reservoir computing},\ }\href@noop {} {\bibfield  {journal} {\bibinfo  {journal} {arXiv preprint arXiv:2302.03595}\ } (\bibinfo {year} {2023})}\BibitemShut {NoStop}%
\bibitem [{\citenamefont {Mackey}\ and\ \citenamefont {Glass}(1977)}]{mackey1977oscillation}%
  \BibitemOpen
  \bibfield  {author} {\bibinfo {author} {\bibfnamefont {M.~C.}\ \bibnamefont {Mackey}}\ and\ \bibinfo {author} {\bibfnamefont {L.}~\bibnamefont {Glass}},\ }\bibfield  {title} {\bibinfo {title} {Oscillation and chaos in physiological control systems},\ }\href@noop {} {\bibfield  {journal} {\bibinfo  {journal} {Science}\ }\textbf {\bibinfo {volume} {197}},\ \bibinfo {pages} {287} (\bibinfo {year} {1977})}\BibitemShut {NoStop}%
\bibitem [{\citenamefont {Lee}\ and\ \citenamefont {Jeong}(2011)}]{lee2011quantification}%
  \BibitemOpen
  \bibfield  {author} {\bibinfo {author} {\bibfnamefont {C.-W.}\ \bibnamefont {Lee}}\ and\ \bibinfo {author} {\bibfnamefont {H.}~\bibnamefont {Jeong}},\ }\bibfield  {title} {\bibinfo {title} {Quantification of macroscopic quantum superpositions within phase space},\ }\href@noop {} {\bibfield  {journal} {\bibinfo  {journal} {Physical review letters}\ }\textbf {\bibinfo {volume} {106}},\ \bibinfo {pages} {220401} (\bibinfo {year} {2011})}\BibitemShut {NoStop}%
\bibitem [{\citenamefont {Gardiner}\ \emph {et~al.}(2004)\citenamefont {Gardiner}, \citenamefont {Zoller},\ and\ \citenamefont {Zoller}}]{gardiner2004quantum}%
  \BibitemOpen
  \bibfield  {author} {\bibinfo {author} {\bibfnamefont {C.}~\bibnamefont {Gardiner}}, \bibinfo {author} {\bibfnamefont {P.}~\bibnamefont {Zoller}},\ and\ \bibinfo {author} {\bibfnamefont {P.}~\bibnamefont {Zoller}},\ }\href@noop {} {\emph {\bibinfo {title} {Quantum noise: a handbook of Markovian and non-Markovian quantum stochastic methods with applications to quantum optics}}}\ (\bibinfo  {publisher} {Springer Science \& Business Media},\ \bibinfo {year} {2004})\BibitemShut {NoStop}%
\bibitem [{\citenamefont {Shalev-Shwartz}\ and\ \citenamefont {Ben-David}(2014)}]{shalev2014understanding}%
  \BibitemOpen
  \bibfield  {author} {\bibinfo {author} {\bibfnamefont {S.}~\bibnamefont {Shalev-Shwartz}}\ and\ \bibinfo {author} {\bibfnamefont {S.}~\bibnamefont {Ben-David}},\ }\href@noop {} {\emph {\bibinfo {title} {Understanding machine learning: From theory to algorithms}}}\ (\bibinfo  {publisher} {Cambridge university press},\ \bibinfo {year} {2014})\BibitemShut {NoStop}%
\bibitem [{\citenamefont {Groisman}\ \emph {et~al.}(2007)\citenamefont {Groisman}, \citenamefont {Kenigsberg},\ and\ \citenamefont {Mor}}]{groisman2007quantumness}%
  \BibitemOpen
  \bibfield  {author} {\bibinfo {author} {\bibfnamefont {B.}~\bibnamefont {Groisman}}, \bibinfo {author} {\bibfnamefont {D.}~\bibnamefont {Kenigsberg}},\ and\ \bibinfo {author} {\bibfnamefont {T.}~\bibnamefont {Mor}},\ }\bibfield  {title} {\bibinfo {title} {" quantumness" versus" classicality" of quantum states},\ }\href@noop {} {\bibfield  {journal} {\bibinfo  {journal} {arXiv preprint quant-ph/0703103}\ } (\bibinfo {year} {2007})}\BibitemShut {NoStop}%
\bibitem [{\citenamefont {Ollivier}\ and\ \citenamefont {Zurek}(2001)}]{ollivier2001quantum}%
  \BibitemOpen
  \bibfield  {author} {\bibinfo {author} {\bibfnamefont {H.}~\bibnamefont {Ollivier}}\ and\ \bibinfo {author} {\bibfnamefont {W.~H.}\ \bibnamefont {Zurek}},\ }\bibfield  {title} {\bibinfo {title} {Quantum discord: a measure of the quantumness of correlations},\ }\href@noop {} {\bibfield  {journal} {\bibinfo  {journal} {Physical review letters}\ }\textbf {\bibinfo {volume} {88}},\ \bibinfo {pages} {017901} (\bibinfo {year} {2001})}\BibitemShut {NoStop}%
\bibitem [{\citenamefont {Takahashi}(1986)}]{takahashi1986wigner}%
  \BibitemOpen
  \bibfield  {author} {\bibinfo {author} {\bibfnamefont {K.}~\bibnamefont {Takahashi}},\ }\bibfield  {title} {\bibinfo {title} {Wigner and husimi functions in quantum mechanics},\ }\href@noop {} {\bibfield  {journal} {\bibinfo  {journal} {Journal of the Physical Society of Japan}\ }\textbf {\bibinfo {volume} {55}},\ \bibinfo {pages} {762} (\bibinfo {year} {1986})}\BibitemShut {NoStop}%
\bibitem [{\citenamefont {Fr{\"o}wis}\ \emph {et~al.}(2018)\citenamefont {Fr{\"o}wis}, \citenamefont {Sekatski}, \citenamefont {D{\"u}r}, \citenamefont {Gisin},\ and\ \citenamefont {Sangouard}}]{frowis2018macroscopic}%
  \BibitemOpen
  \bibfield  {author} {\bibinfo {author} {\bibfnamefont {F.}~\bibnamefont {Fr{\"o}wis}}, \bibinfo {author} {\bibfnamefont {P.}~\bibnamefont {Sekatski}}, \bibinfo {author} {\bibfnamefont {W.}~\bibnamefont {D{\"u}r}}, \bibinfo {author} {\bibfnamefont {N.}~\bibnamefont {Gisin}},\ and\ \bibinfo {author} {\bibfnamefont {N.}~\bibnamefont {Sangouard}},\ }\bibfield  {title} {\bibinfo {title} {Macroscopic quantum states: Measures, fragility, and implementations},\ }\href@noop {} {\bibfield  {journal} {\bibinfo  {journal} {Reviews of Modern Physics}\ }\textbf {\bibinfo {volume} {90}},\ \bibinfo {pages} {025004} (\bibinfo {year} {2018})}\BibitemShut {NoStop}%
\bibitem [{\citenamefont {Nimmrichter}\ and\ \citenamefont {Hornberger}(2013)}]{nimmrichter2013macroscopicity}%
  \BibitemOpen
  \bibfield  {author} {\bibinfo {author} {\bibfnamefont {S.}~\bibnamefont {Nimmrichter}}\ and\ \bibinfo {author} {\bibfnamefont {K.}~\bibnamefont {Hornberger}},\ }\bibfield  {title} {\bibinfo {title} {Macroscopicity of mechanical quantum superposition states},\ }\href@noop {} {\bibfield  {journal} {\bibinfo  {journal} {Physical review letters}\ }\textbf {\bibinfo {volume} {110}},\ \bibinfo {pages} {160403} (\bibinfo {year} {2013})}\BibitemShut {NoStop}%
\bibitem [{\citenamefont {Leggett}\ and\ \citenamefont {Garg}(1985)}]{leggett1985quantum}%
  \BibitemOpen
  \bibfield  {author} {\bibinfo {author} {\bibfnamefont {A.~J.}\ \bibnamefont {Leggett}}\ and\ \bibinfo {author} {\bibfnamefont {A.}~\bibnamefont {Garg}},\ }\bibfield  {title} {\bibinfo {title} {Quantum mechanics versus macroscopic realism: Is the flux there when nobody looks?},\ }\href@noop {} {\bibfield  {journal} {\bibinfo  {journal} {Physical Review Letters}\ }\textbf {\bibinfo {volume} {54}},\ \bibinfo {pages} {857} (\bibinfo {year} {1985})}\BibitemShut {NoStop}%
\bibitem [{\citenamefont {Leggett}(2016)}]{leggett2016note}%
  \BibitemOpen
  \bibfield  {author} {\bibinfo {author} {\bibfnamefont {A.~J.}\ \bibnamefont {Leggett}},\ }\bibfield  {title} {\bibinfo {title} {Note on the" size" of schroedinger cats},\ }\href@noop {} {\bibfield  {journal} {\bibinfo  {journal} {arXiv preprint arXiv:1603.03992}\ } (\bibinfo {year} {2016})}\BibitemShut {NoStop}%
\bibitem [{\citenamefont {Gong}(2011)}]{I}%
  \BibitemOpen
  \bibfield  {author} {\bibinfo {author} {\bibfnamefont {J.}~\bibnamefont {Gong}},\ }\href {https://doi.org/10.48550/ARXIV.1106.0062} {\bibinfo {title} {Comment on "quantification of macroscopic quantum superpositions within phase space"}} (\bibinfo {year} {2011})\BibitemShut {NoStop}%
\bibitem [{\citenamefont {Haar}(1933)}]{haar1933massbegriff}%
  \BibitemOpen
  \bibfield  {author} {\bibinfo {author} {\bibfnamefont {A.}~\bibnamefont {Haar}},\ }\bibfield  {title} {\bibinfo {title} {Der massbegriff in der theorie der kontinuierlichen gruppen},\ }\href@noop {} {\bibfield  {journal} {\bibinfo  {journal} {Annals of mathematics}\ ,\ \bibinfo {pages} {147}} (\bibinfo {year} {1933})}\BibitemShut {NoStop}%
\bibitem [{\citenamefont {Watrous}(2018)}]{watrous2018theory}%
  \BibitemOpen
  \bibfield  {author} {\bibinfo {author} {\bibfnamefont {J.}~\bibnamefont {Watrous}},\ }\href@noop {} {\emph {\bibinfo {title} {The theory of quantum information}}}\ (\bibinfo  {publisher} {Cambridge university press},\ \bibinfo {year} {2018})\BibitemShut {NoStop}%
\bibitem [{\citenamefont {Steuernagel}\ and\ \citenamefont {Lee}(2023)}]{steuernagel2023quantumness}%
  \BibitemOpen
  \bibfield  {author} {\bibinfo {author} {\bibfnamefont {O.}~\bibnamefont {Steuernagel}}\ and\ \bibinfo {author} {\bibfnamefont {R.-K.}\ \bibnamefont {Lee}},\ }\bibfield  {title} {\bibinfo {title} {Quantumness measure from phase space distributions},\ }\href@noop {} {\bibfield  {journal} {\bibinfo  {journal} {arXiv preprint arXiv:2311.17399}\ } (\bibinfo {year} {2023})}\BibitemShut {NoStop}%
\bibitem [{\citenamefont {Bertet}\ \emph {et~al.}(2012)\citenamefont {Bertet}, \citenamefont {Ong}, \citenamefont {Boissonneault}, \citenamefont {Bolduc}, \citenamefont {Mallet}, \citenamefont {Doherty}, \citenamefont {Blais}, \citenamefont {Vion},\ and\ \citenamefont {Esteve}}]{bertet2012circuit}%
  \BibitemOpen
  \bibfield  {author} {\bibinfo {author} {\bibfnamefont {P.}~\bibnamefont {Bertet}}, \bibinfo {author} {\bibfnamefont {F.}~\bibnamefont {Ong}}, \bibinfo {author} {\bibfnamefont {M.}~\bibnamefont {Boissonneault}}, \bibinfo {author} {\bibfnamefont {A.}~\bibnamefont {Bolduc}}, \bibinfo {author} {\bibfnamefont {F.}~\bibnamefont {Mallet}}, \bibinfo {author} {\bibfnamefont {A.}~\bibnamefont {Doherty}}, \bibinfo {author} {\bibfnamefont {A.}~\bibnamefont {Blais}}, \bibinfo {author} {\bibfnamefont {D.}~\bibnamefont {Vion}},\ and\ \bibinfo {author} {\bibfnamefont {D.}~\bibnamefont {Esteve}},\ }\href@noop {} {\emph {\bibinfo {title} {Circuit quantum electrodynamics with a nonlinear resonator}}}\ (\bibinfo  {publisher} {Oxford University Press},\ \bibinfo {year} {2012})\BibitemShut {NoStop}%
\bibitem [{\citenamefont {R{\"o}ssler}(1976)}]{rossler1976equation}%
  \BibitemOpen
  \bibfield  {author} {\bibinfo {author} {\bibfnamefont {O.~E.}\ \bibnamefont {R{\"o}ssler}},\ }\bibfield  {title} {\bibinfo {title} {An equation for continuous chaos},\ }\href@noop {} {\bibfield  {journal} {\bibinfo  {journal} {Physics Letters A}\ }\textbf {\bibinfo {volume} {57}},\ \bibinfo {pages} {397} (\bibinfo {year} {1976})}\BibitemShut {NoStop}%
\bibitem [{\citenamefont {Rossler}(1979)}]{rossler1979equation}%
  \BibitemOpen
  \bibfield  {author} {\bibinfo {author} {\bibfnamefont {O.}~\bibnamefont {Rossler}},\ }\bibfield  {title} {\bibinfo {title} {An equation for hyperchaos},\ }\href@noop {} {\bibfield  {journal} {\bibinfo  {journal} {Physics Letters A}\ }\textbf {\bibinfo {volume} {71}},\ \bibinfo {pages} {155} (\bibinfo {year} {1979})}\BibitemShut {NoStop}%
\bibitem [{\citenamefont {Kubota}\ \emph {et~al.}(2022)\citenamefont {Kubota}, \citenamefont {Suzuki}, \citenamefont {Kobayashi}, \citenamefont {Tran}, \citenamefont {Yamamoto},\ and\ \citenamefont {Nakajima}}]{noise}%
  \BibitemOpen
  \bibfield  {author} {\bibinfo {author} {\bibfnamefont {T.}~\bibnamefont {Kubota}}, \bibinfo {author} {\bibfnamefont {Y.}~\bibnamefont {Suzuki}}, \bibinfo {author} {\bibfnamefont {S.}~\bibnamefont {Kobayashi}}, \bibinfo {author} {\bibfnamefont {Q.~H.}\ \bibnamefont {Tran}}, \bibinfo {author} {\bibfnamefont {N.}~\bibnamefont {Yamamoto}},\ and\ \bibinfo {author} {\bibfnamefont {K.}~\bibnamefont {Nakajima}},\ }\href {https://doi.org/10.48550/ARXIV.2207.07924} {\bibinfo {title} {Quantum noise-induced reservoir computing}} (\bibinfo {year} {2022})\BibitemShut {NoStop}%
\bibitem [{\citenamefont {Fry}\ \emph {et~al.}(2023)\citenamefont {Fry}, \citenamefont {Deshmukh}, \citenamefont {Chen}, \citenamefont {Rastunkov},\ and\ \citenamefont {Markov}}]{fry2023optimizing}%
  \BibitemOpen
  \bibfield  {author} {\bibinfo {author} {\bibfnamefont {D.}~\bibnamefont {Fry}}, \bibinfo {author} {\bibfnamefont {A.}~\bibnamefont {Deshmukh}}, \bibinfo {author} {\bibfnamefont {S.~Y.-C.}\ \bibnamefont {Chen}}, \bibinfo {author} {\bibfnamefont {V.}~\bibnamefont {Rastunkov}},\ and\ \bibinfo {author} {\bibfnamefont {V.}~\bibnamefont {Markov}},\ }\bibfield  {title} {\bibinfo {title} {Optimizing quantum noise-induced reservoir computing for nonlinear and chaotic time series prediction},\ }\href@noop {} {\bibfield  {journal} {\bibinfo  {journal} {arXiv preprint arXiv:2303.05488}\ } (\bibinfo {year} {2023})}\BibitemShut {NoStop}%
\end{thebibliography}%

\clearpage

\widetext
\begin{center}
\textbf{\large Supplementary Material for: Correlations Between Quantumness and Learning Performance in Reservoir Computing with a Single Oscillator}
\end{center}
\setcounter{equation}{0}
\setcounter{figure}{0}
\setcounter{table}{0}
\setcounter{page}{1}
\setcounter{section}{0}
\makeatletter
\renewcommand{\theequation}{S\arabic{equation}}
\renewcommand{\thefigure}{S\arabic{figure}}
\renewcommand{\bibnumfmt}[1]{[S#1]}
\renewcommand{\citenumfont}[1]{S#1}

\section{R\"ossler Attractor}\label{sec:chaos}
\begin{figure}[h]
    \centering
    \subfloat[The oscillator is trained to reproduce the R\"ossler time series, and then asked to perform prediction, given initial values, which are the data points before the dashed line. The parameters are $(\alpha,\kappa,K) = (1.0, 0.2, 0.01)$.]{\includegraphics[width=0.4\textwidth]{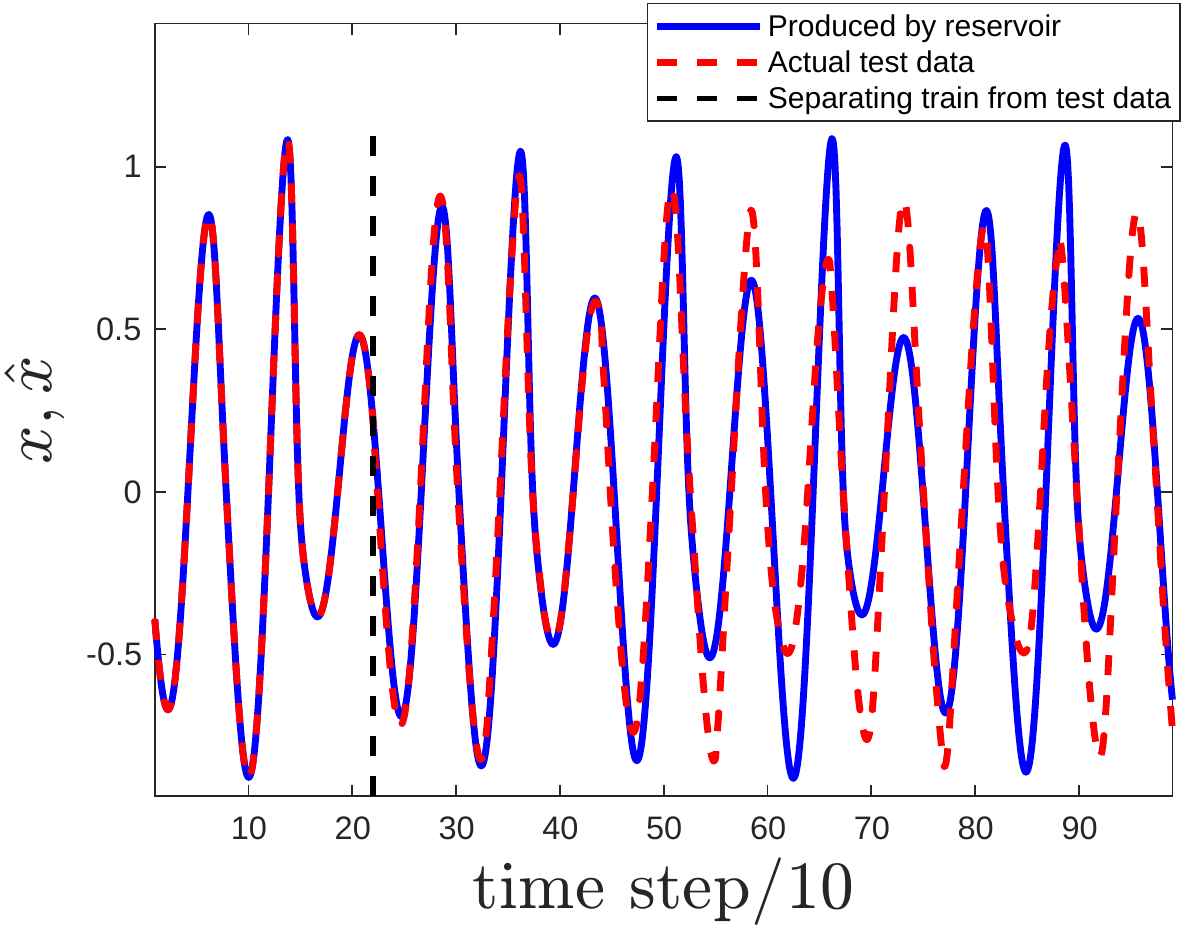}\label{fig:Ros1}}
    \hspace{1 cm}
  \subfloat[Comparing the phase diagrams $X-Y$ of the R\"ossler attractor of the actual R\"ossler attractor with the predicted by the reservoir (left) with that of the reservoir's output (right).]{\includegraphics[width=0.4\textwidth]{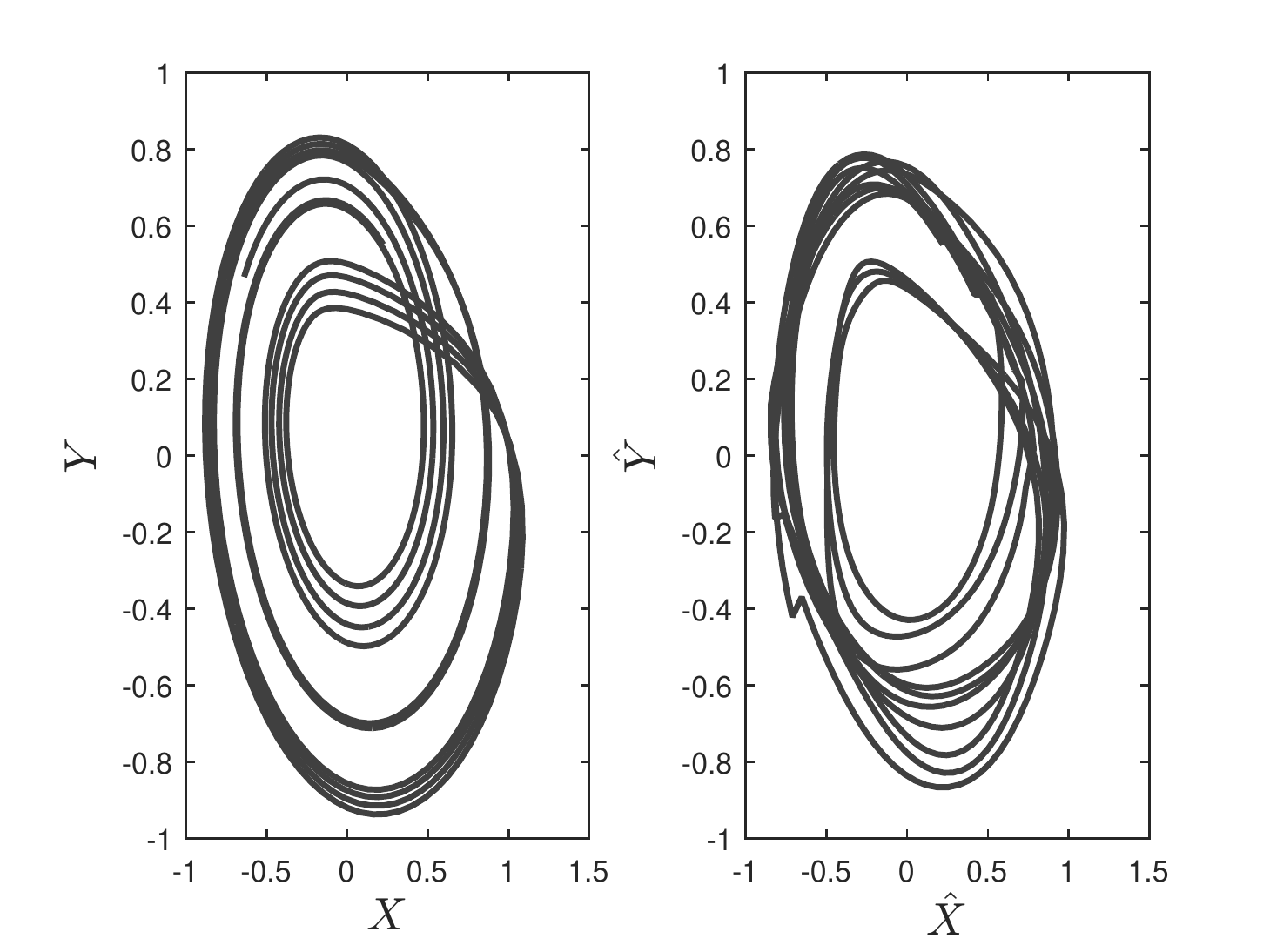}\label{fig:Ros2}}
  \caption{R\"ossler attractor training with a quantum oscillator. Here $n$ denotes the time step or equivalently the index of the sampled point from the function, $(x,y)$ and $(\hat{x},\hat{y})$ refer to the actual and the predicted series, respectively.}
  \label{fig:Rossler}
\end{figure}

This section provides the performance in the training of R\"ossler attractor. The R\"ossler attractor \cite{rossler1976equation, rossler1979equation} is a three-dimensional motion following the dynamics
\begin{equation}
    \begin{cases}
    \frac{dx}{dt} &= -y-z\\
    \frac{dy}{dt} &= x + ay\\
    \frac{dz}{dt} &= b+ z(x-c)
    \end{cases}
\end{equation}
where $(a,b,c)\in\mathbb R^3$ are constant parameters, which we set to $(0.2,0.2,5.7)$ in our experiment. \cref{fig:Ros1} and \cref{fig:Ros2} show the model's learning results for this chaotic time series. We highlight that each component of the oscillator is learned independently from the others.

\section{Noise}\label{SM:noise}
Noise is an inevitable factor in quantum devices, and it is of profound importance for a quantum computing approach to be robust to noise. We show the robustness of this approach by considering different noise models as explained below.

\begin{figure}[h]
    \centering
    \subfloat[Prediction of Mackey-Glass under the effects of noise.
The noise models considered are dephasing incoherent
pumping, and white noise on the input.]{
    \includegraphics[width=0.45\textwidth]{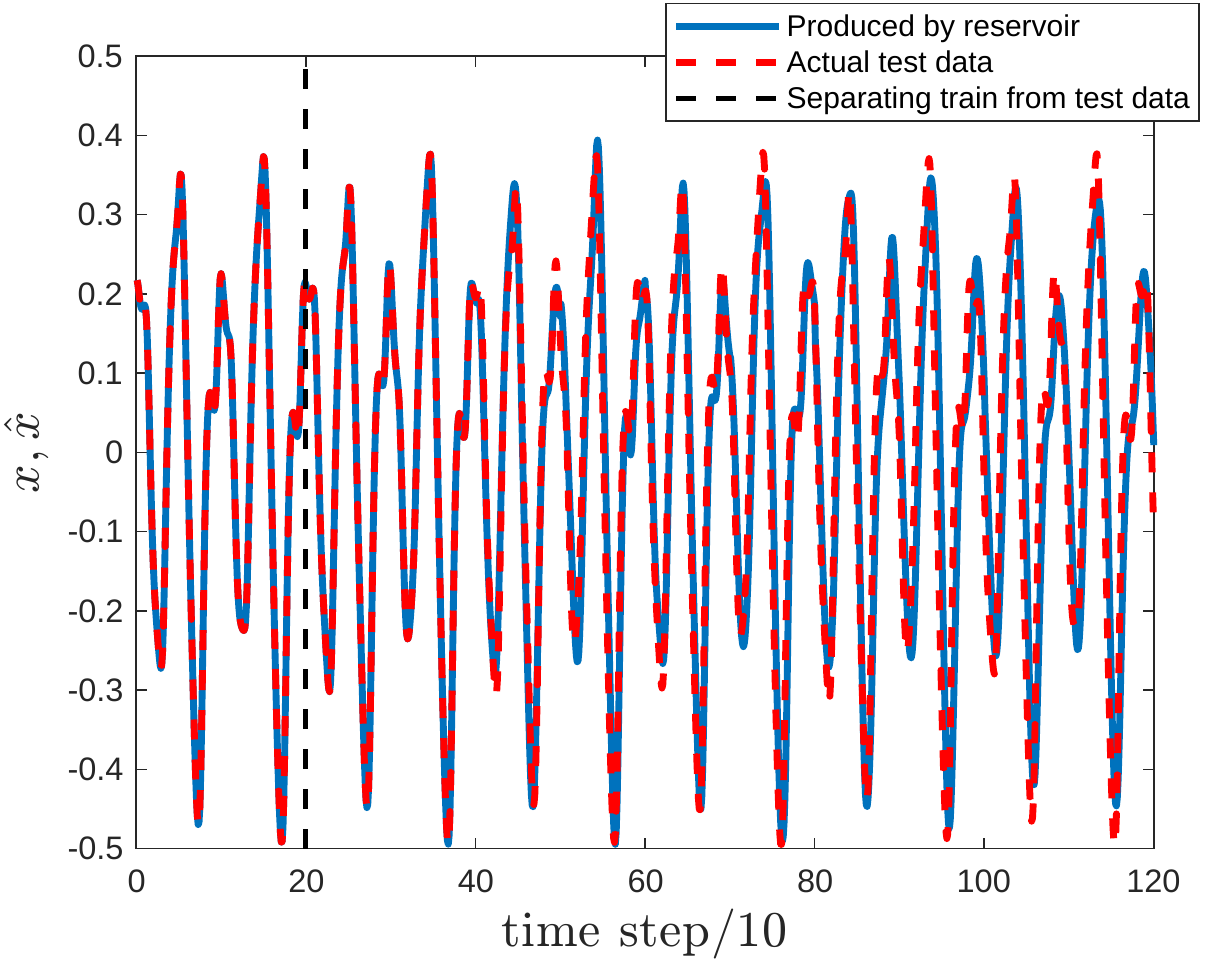}\label{fig:noisepred}
    }
    \hfill
    \subfloat[Comparing the delayed embedding of the actual
Mackey-Glass (left) with the reservoir’s output (right) in
the noisy settings.]{
    \includegraphics[width=0.45\textwidth]{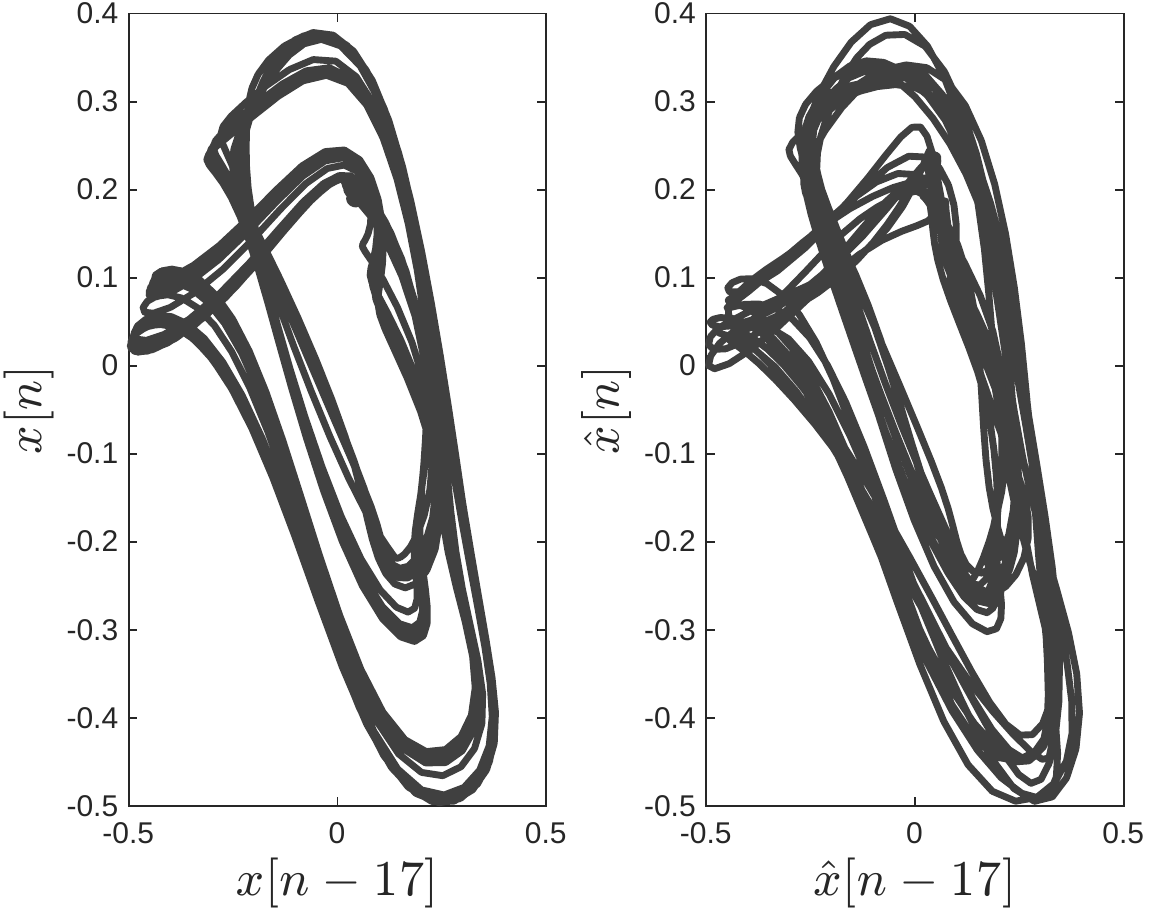}\label{fig:noisephase}}
    
    \caption{Noisy reservoir learning MG. The MG training process is performed when there are a variety of noises applied to the reservoir.}
    \label{fig:NoisyMG}
\end{figure}

\paragraph{Noisy MG learning} Here we consider adding noise to the learning process of an MG time series, and will examine the performance. We introduce a dephasing error by considering the Lindbladian operators $L_n = \lambda \ket{n}\bra{n}$ for all $n\in \mathbb Z_{\geq 0}$. Furthermore, the incoherent pumping error is simulated by considering the Lindbladian corresponding to $\lambda a^\dagger$. On top of those, we add white noise to the input of the reservoir. This is performed via changing the equations of evolution \eqref{eq:ev-2} through the substitution $f(t) \rightarrow f(t) + \lambda' n(t)$. Here, $n(t)$ is the white noise of unit power, and $\lambda'$ controls its strength. \cref{fig:NoisyMG} shows the performance of the reservoir's output under both incoherent pumping, dephasing error, and white noise (see \cref{eq:ev-2}). We have set $(K,\kappa, \alpha, \lambda, \lambda')= (0.05, 0.15, 1.2, 0.05, 0.02)$.

\begin{figure}[h]
    \centering
    \includegraphics[width = 0.4\textwidth]{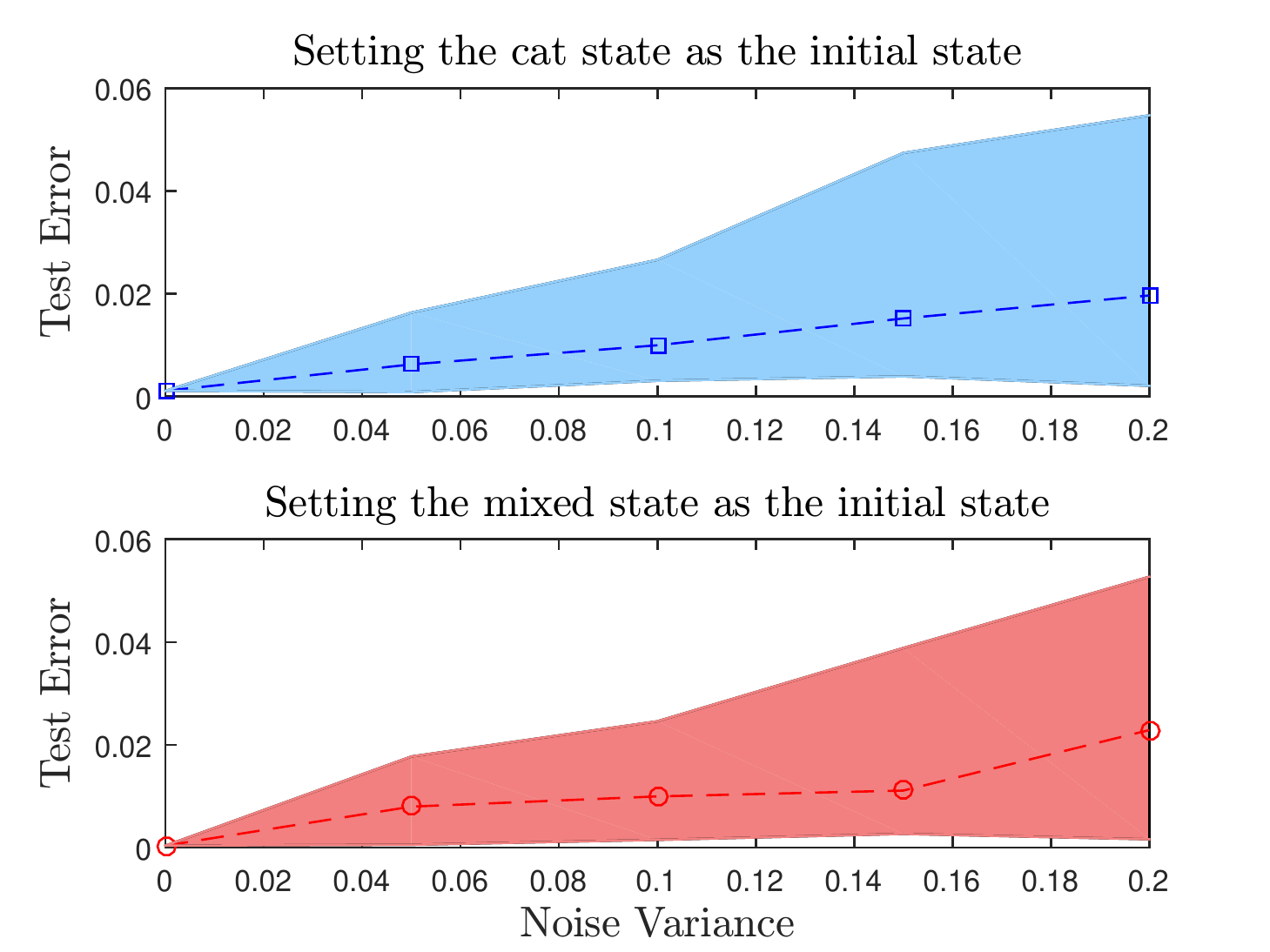}
    \caption{Test error of training noisy sawtooth function. The initial states are the cat state and its classical mixture i.e., the normalized $\ket{\alpha} + \ket{-\alpha}$ and $\ket{\alpha}\bra\alpha + \ket{-\alpha}\bra{-\alpha}$. The noise model in this example is an additive white noise to the input.}
    \label{fig:stn}
\end{figure}

\begin{figure}[h]
    \centering
    \includegraphics[width = 0.5\textwidth]{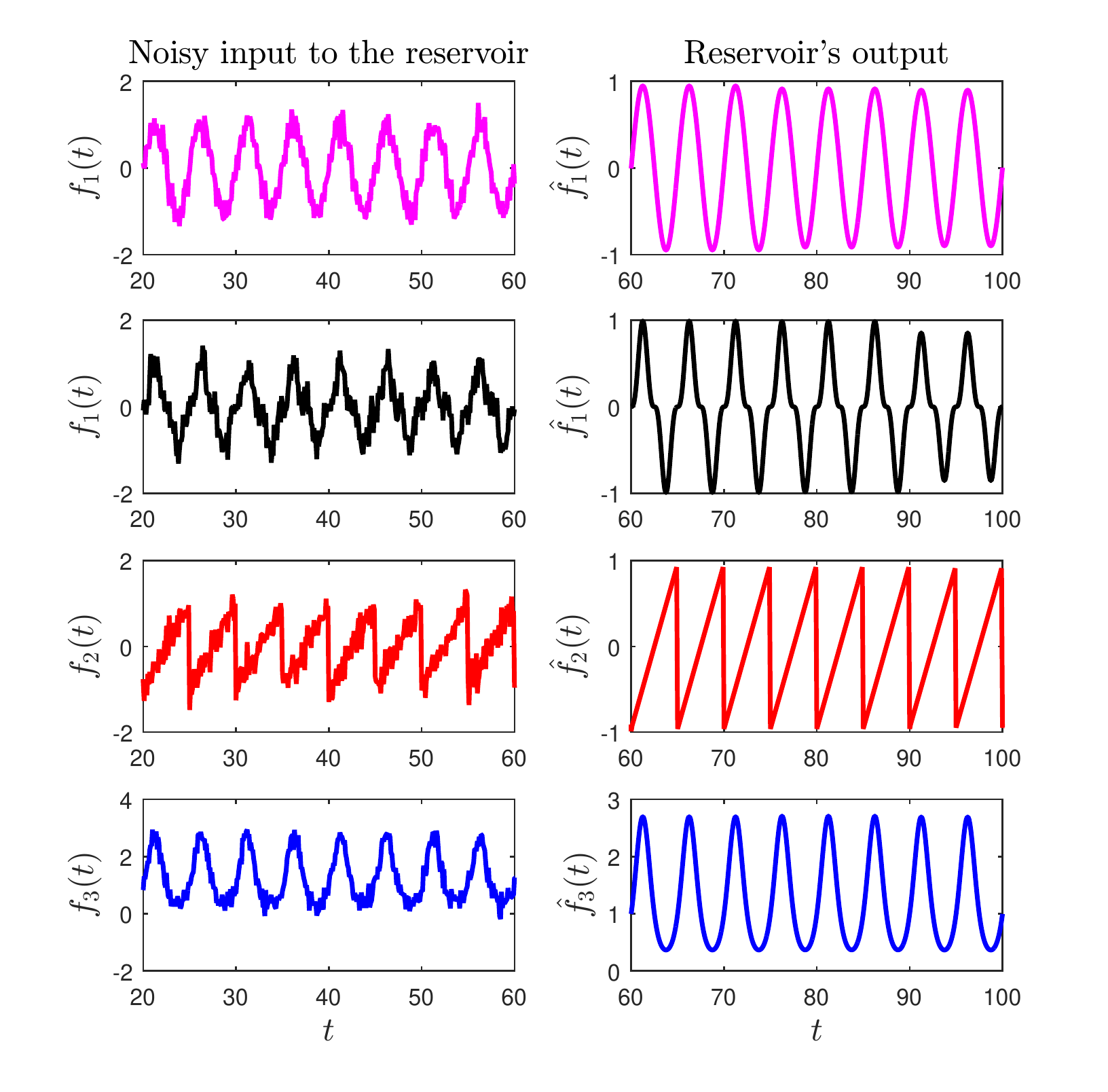}
    \caption{Learning noisy periodic functions. The input to the reservoir is contaminated by white noise. However, the reservoir is still able to learn the input signal.}
    \label{fig:ArbF}
\end{figure}

\paragraph{Other time series} Here we consider the effect of noise on several other time series. We consier simple periodic functions, but we add a much larger white noise to the input. \cref{fig:ArbF} shows the outcome of this learning task. Despite significant signal distortion caused by noise, the oscillator demonstrates the ability to learn the underlying periodic functions. We further investigated the effect of training sawtooth signal with different noise levels, which resulted in \cref{fig:stn}. A similar experiment, this time with the MG, resulted in a training error of $0.053$. Noting that the training error in the noiseless case results in an error of $0.047$, we conclude that the model is robust to this noise model for a variety of prediction tasks. We made the choice of parameters $\alpha = 0.1$, $\kappa = K = 0.05$ in obtaining the results.

It is worth mentioning that noise in the context of reservoir computing is shown to be useful in certain cases
\cite{noise, fry2023optimizing}.

\section{Mackey--Glass with different parameters}\label{app:changing-tau}

In this section we investigate the hardness of training a Mackey--Glass series with different $\tau$ values. Note that $\tau$ significanly changes the structure of the chaotic series. A small $\tau$ corresponds to a simple series, while a large $\tau$ bring in much chaos. We trained our model on series with $\tau\in\{10,17,40\}$ and provided the traninig result in . We highlight that all hyperparameters used in the training of these series are the same except for the input length, $N$ (as defined in \cref{sec:rc}). We have used $N=110$, $N=200$, and $N=400$ for $\tau=10$, $\tau=17$, and $\tau=40$, respectively. Notably, using an input length as large as $N=200$ results in the poor training for the series with $\tau=40$, while an input length as small as $N=110$ is sufficient for learning the series with $\tau=10$.

\begin{figure}
    \centering
    \includegraphics[width=0.5\textwidth]{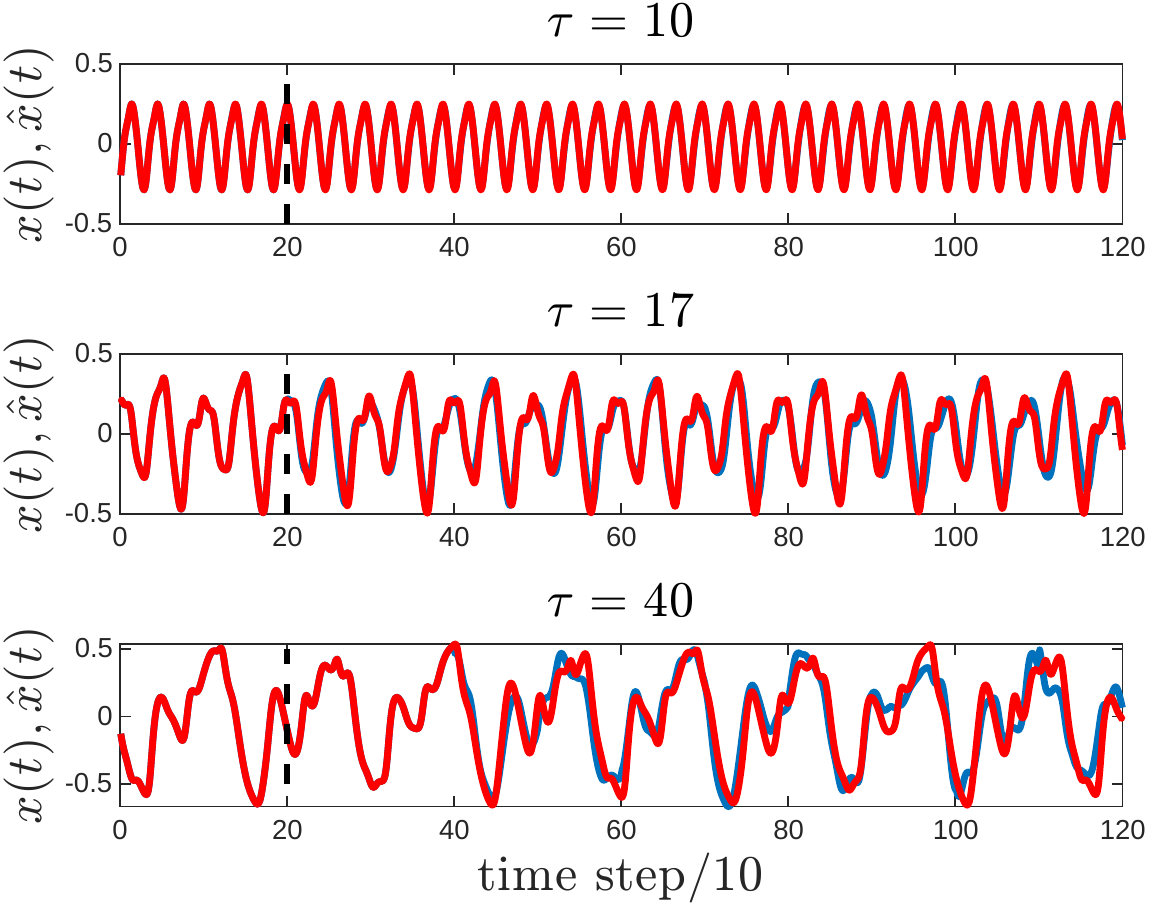}
    \caption{Training MG series with different $\tau$ values. We have used the values $\tau\in\{10,17,40\}$. The blue curve in each panel demonstrates the output of the reservoir and the red curve corresponds to the actual data. As we observe, the behavior of the series get more complex and chaotic as we increase $\tau$. We also highlight that all the hyperparameters of the reservoirs we used in this simulation are the same, except for the training length $N$ (as defined in \cref{sec:rc}). We use $N=110,200,400$ for the generation of the top, middle, and the bottom panel respectively.}
    \label{fig:enter-label}
\end{figure}

\section{Random states used in Section \ref{sec:ParamSpace}}\label{sec:random-states}
\begin{figure}[h]
    \centering
 \includegraphics[width=0.7\textwidth]{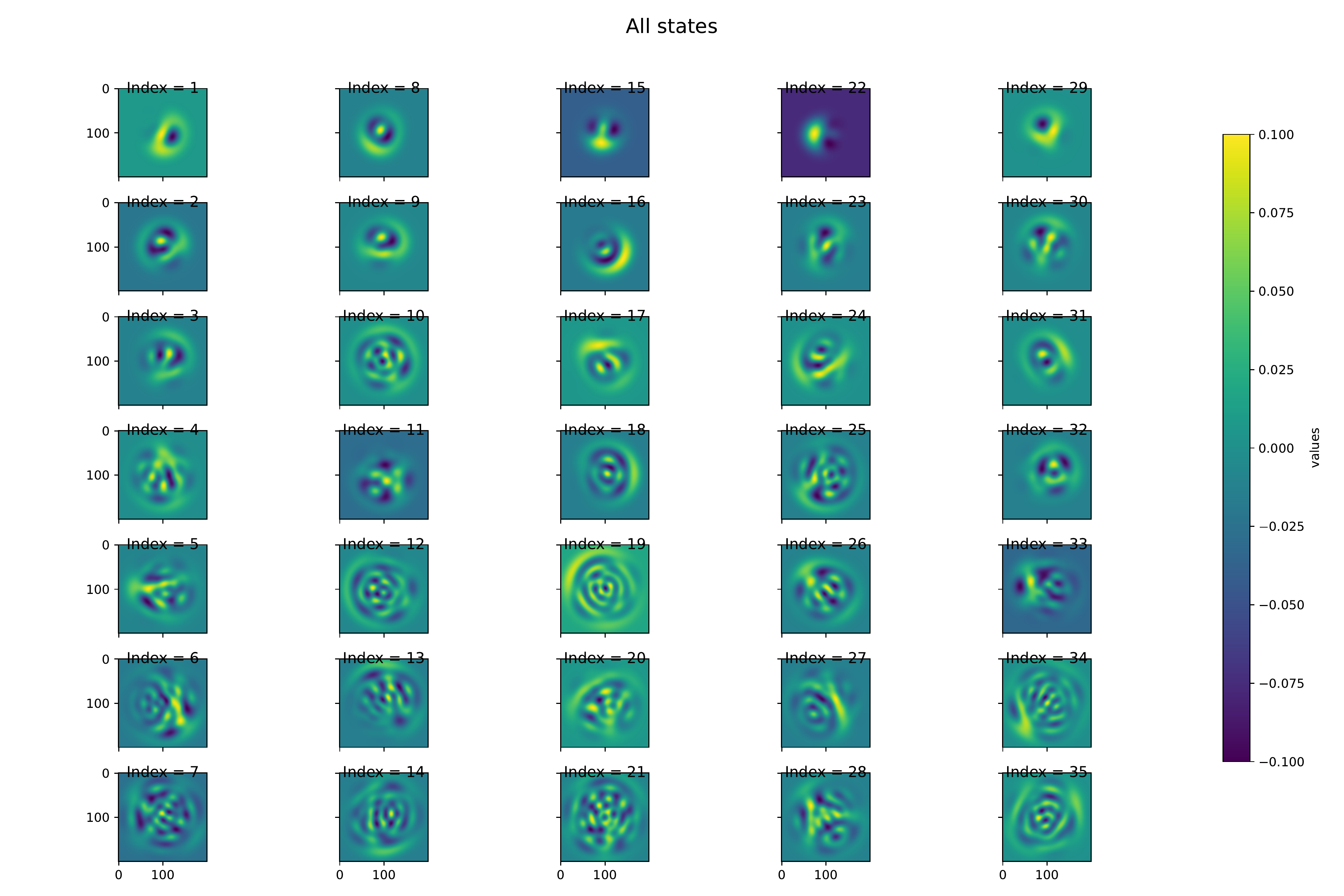}
    \caption{The Wigner function of all the $35$ randomly selected states that were used in the training process described in \cref{sec:ParamSpace}.}
    \label{fig:AllStates}
\end{figure}

\begin{figure}[h]
    \centering
    \includegraphics[width=.45\textwidth]{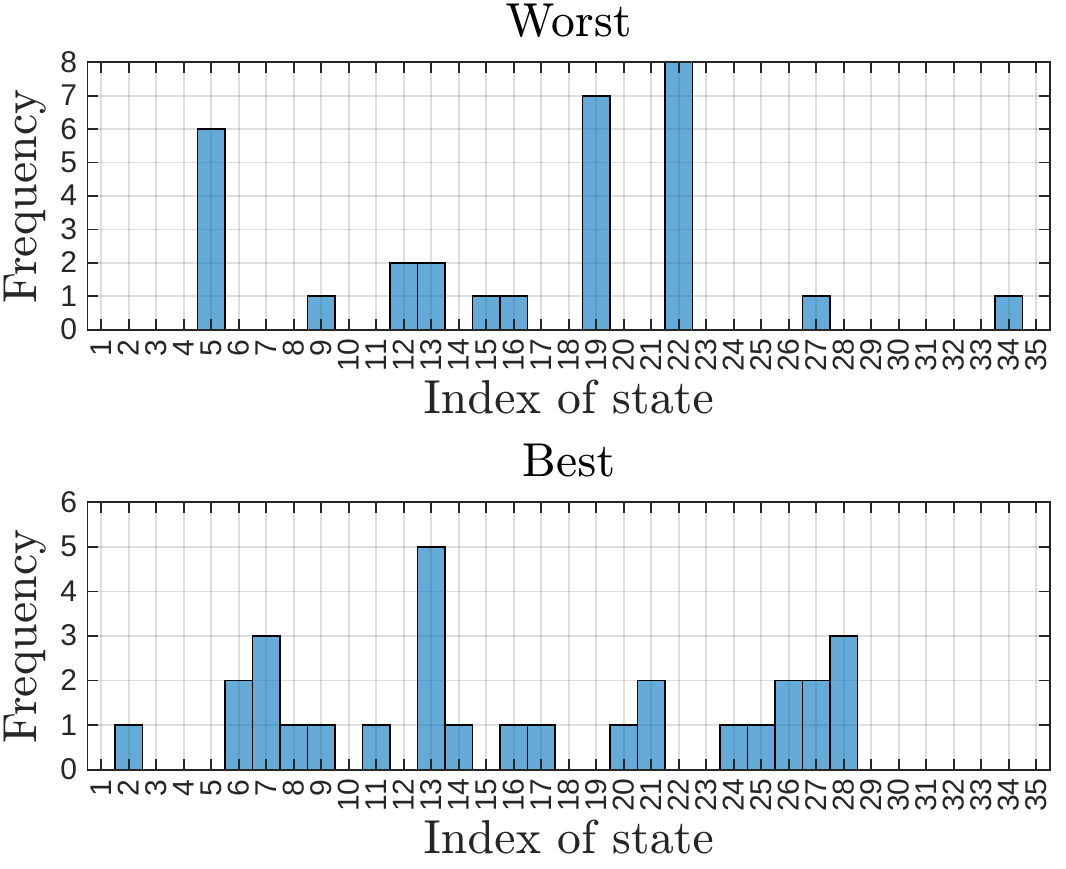}
    \caption{The histogram that shows how many times (i.e., for how many of $(K,\kappa)$ pairs) does a state appears as the worst (top panel) and the best (bottom panel) performing state. Please refer to \cref{fig:AllStates}, where we represent the Wigner function of these randomly selected states.}
    \label{fig:GoodNBad}
\end{figure}

We have presented the Wigner function of the $35$ randomly selected states that are sent through the training process in \cref{sec:ParamSpace}, in \cref{fig:AllStates}. We have also explored the performance of the state in \cref{fig:GoodNBad}. Notably, states indexed as 22, 19, and 5 should be avoided during training, as they consistently rank among the worst performers and never appear among the best ones in \cref{fig:GoodNBad}. However, it is noteworthy that the best-performing states are dispersed and are not well-concentrated on a few instances.

\section{Evolution Animations}\label{sec:evolution-animation}
Animations showing the evolution of the Wigner function throughout the process are prepared and made available online at \href{https://github.com/arsalan-motamedi/QRC/tree/main/EvolutionAnimations}{https://github.com/arsalan-motamedi/QRC/tree/main/EvolutionAnimations}.

\end{document}